\documentclass[preprint,twocolumn]{aastex631}
\usepackage{CJKutf8}
\usepackage{amsmath}

\shorttitle{}
\shortauthors{}

\graphicspath{{./}{figures/}}

\begin{document}
\begin{CJK*}{UTF8}{gbsn}

\title[Enhanced rate of TDEs due to SSR]{Enhancement of the Rate of Tidal Disruption Events in Active Galactic Nuclei due to the Sweeping Secular Resonance Mechanism}

\author[0000-0002-7814-9185]{Xiaochen Zheng（郑晓晨)}
\affiliation{Beijing Planetarium, Beijing Academy of Science and Technology \\
No. 138 Xizhimenwai Main Street, Beijing 100044, China}

\author[0000-0002-1417-8024]{Morgan MacLeod}
\affiliation{Harvard-Smithsonian Center for Astrophysics \\
60 Garden Street, Cambridge, MA, 02138, USA}

\author[0000-0001-5466-4628]{Douglas N. C. Lin（林潮）}
\affiliation{Department of Astronomy and Astrophysics, University of California Santa Cruz \\
CA 95064, USA}
\affiliation{Institute for Advanced Studies, Tsinghua University \\
Beijing 100084, China}

\author{Yi Yang (杨宜)}
\affiliation{Beijing Planetarium, Beijing Academy of Science and Technology \\
No. 138 Xizhimenwai Main Street, Beijing 100044, China}

\author[0000-0002-4448-3679]{Zhenzhen Shao (邵珍珍)}
\affiliation{Beijing Planetarium, Beijing Academy of Science and Technology \\
No. 138 Xizhimenwai Main Street, Beijing 100044, China}

\correspondingauthor{Xiaochen Zheng, Morgan MacLeod, Douglas N. C. Lin}
\email{x.c.zheng1989@gmail.com, morgan.macleod@cfa.harvard.edu, lin@ucolick.org}

% These dates will be filled out by the publisher
%\date{Accepted XXX. Received YYY; in original form ZZZ}

% Enter the current year, for the copyright statements etc.
%\pubyear{2020}

% Don't change these lines
%\begin{document}
%\label{firstpage}
%\pagerange{\pageref{firstpage}--\pageref{lastpage}}
%\maketitle

% Abstract of the paper
\begin{abstract}
Tidal disruption event (TDE) rates in active galactic nuclei (AGN) consistently exceed predictions from two-body relaxation, particularly in post-starburst and green valley galaxies. We explain this excess with a new mechanism: a sweeping secular resonance (SSR) driven by an intermediate-mass companion (IMC) and a depleting gaseous disk. As the disk mass declines, a resonance between stellar and IMC orbital precession sweeps through the nuclear cluster, exciting stellar eccentricities to near unity on orbital timescales far faster than gravitational relaxation.
Our analytical framework, validated by N-body simulations (REBOUND), shows this mechanism requires IMC-to-SMBH mass ratios of $q \geq 10^{-3}$, disk mass ratio $p \geq 10^{-3}$, and few Myr-scale disk depletion. It is highly effective for co-orbiting IMCs but negligible for counter-orbiting ones. The TDE rate peaks at $10^{-3}-10^{-2}$
per galaxy per year for a depletion timescale $\tau_{\rm dep} \sim 10$ Myr. Even lower-mass IMCs can produce significant enhancements with compact, long-lived disks.
Our model naturally explains elevated AGN TDE rates and implies that a high TDE incidence is a potential tracer of hidden parsec-scale IMCs, offering testable predictions for future AGN monitoring.
\end{abstract}

%%%%%%%%%%%%%%%%%%%%%%%%%%%%%%%%%%%%%%%%%%%%%%%%%%

%%%%%%%%%%%%%%%%% BODY OF PAPER %%%%%%%%%%%%%%%%%%

\section{Introduction}

The crowded and complex environments of galactic centers are host to dense stellar clusters and central supermassive black holes (SMBHs)
\citep{lynden_bell_galactic_1969, lynden_bell_quasars_1971, kormendy_inward_1995, becklin_infrared_1968, rieke_stellar_1988, genzel_physical_1987, genzel_nucleus_1994, Mezger1996}. 
They intermittently transform into active galactic nuclei (AGNs) when the SMBHs are fed by active accretion-disk flows \citep{lynden_bell_galactic_1969, 
Rees1984}.  These disks are fertile sites for star formation \citep{Paczynski1978, Kolykhalov1980, Lin1987, Shlosman1989, Collin1999, Goodman2003}, capture \citep{Artymowicz1993, Wang2024}, rejuvenation \citep{Davies2020}, and evolution \citep{Cantiello2021}. The density of these environments and the coexistence of these various components set the stage for a range of physical processes that result from interactions between the stars, the SMBH, and a gaseous accretion disk \citep[e.g.][]{Linial2024, Liu2025}. This paper focuses on one such category of interaction. 

Tidal disruption events (TDEs) occur when stars are gravitationally disrupted by close passages by a supermassive black hole (SMBH). The critical impact parameter for a star could be tidally disrupted by an SMBH is called the tidal disruption radius, $R_{t}$, which is given by $R_{t} = R_{\star}(M_{\bullet}/M_{\star})^{1/3}$, for a star of mass $M_{\star}$ and radius $R_{\star}$ around a black hole of mass $M_{\bullet}$.  In the aftermath of a TDE, the former star is stretched into a stream of debris on a range of orbital trajectories \citep{Rees1988,1994ApJ...422..508K}. Some of the debris remains bound to the SMBH, and as it falls back it may assemble into an accretion disk \citep[e.g.][]{2014ApJ...783...23G}, driving an accretion episode accompanied by a potentially observable transient flare \citep{Rees1988,2009MNRAS.392..332L,2013ApJ...767...25G,2021ApJ...906..101M}. 

Observations of TDEs can provide valuable information about the SMBHs responsible for the disruptions \citep[e.g.][]{2019ApJ...872..151M,2020ApJ...904...73R,2021ApJ...907...77Z}, and the stellar populations that surround them \citep[e.g.][]{1999ApJ...514..180U,2012ApJ...757..134M,2013ApJ...777..133M,2014ApJ...794....9M,2017ApJ...841..132L,2016MNRAS.461..371K}. TDEs may also serve as probes of the poorly understood, low-mass end of the SMBH mass function, because stellar dynamical arguments suggest that disruptions are expected to occur at higher rates around lower-mass SMBHs \citep{Magorrian1999}. In principle, the TDE rate depends sensitively on the structure and relaxation processes taking place in the central region of galaxies \citep{2013degn.book.....M}, therefore, the rate of the TDEs can provide information about the SMBH and the stellar distribution in the centre of galaxies \citep[for example, as recently reviewed by][]{2020SSRv..216...35S}.

The signatures of TDEs have been observed across the electromagnetic spectrum, with events being discovered from radio to x-ray wavelengths \citep[e.g.][]{2017ApJ...838..149A, 2020SSRv..216...85S, 2020SSRv..216...81A}. Recent surveys are driving incredible expansions in the number of TDE candidate transients \citep[e.g.][]{2021ApJ...908....4V, Gezari2021, Sazonov2021, Goodwin2022, Angus2022, Hammerstein2023, Jiang2023, Yao2023, Masterson2024, SomalwarRaviLu2025, SomalwarRaviDong2025,  Karmen2025}. Efforts to follow up TDEs across the spectrum reveal that some events are panchromatic, while others are primarily optical or ultraviolet and lack x-ray and radio signatures \citep{2020SSRv..216..124V,2020SSRv..216...85S,2020SSRv..216...81A, 2021ARA&A..59...21G, 2022ApJ...930L...4W, 2022ApJ...939L..33L, 2023Sci...380..656L, 2024ApJ...964L..22H, 2024ApJ...971..185C, 2024ApJ...976...34Y, 2024A&A...692A.262S, 2025A&A...704A...3K, 2025ApJ...993..198Y}, with some combination of a multi-component gaseous flow and a variety of viewing angles likely modulating some of these properties \citep[e.g.][]{2020SSRv..216..114R,2021arXiv210105195L,2022MNRAS.510.5426P, 2022ApJ...937L..28T, 2023MNRAS.518..847G, 2023MNRAS.521.4180B, 2024ApJ...975...94T, 2025arXiv250612729A}. 

As a population of TDE candidates has emerged, a particularly puzzling feature is a very high specific abundance of TDEs in ``green valley" or ``post-starburst" host galaxies \citep[enhanced by a factor of 10 to 100,][]{2020SSRv..216...32F,2021ApJ...908L..20H, Yao2023, 2023ApJ...942....9H, 2025ApJ...989...49X}. The detection of a candidate TDE by \citet{Tadhunter2017} in a sample of 15 nearby ultra-luminous infrared galaxies over a period of just 10 years, which suggested a even higher TDEs rate as $\sim 10^{-2}-10^{-1} {\rm yr}^{-1}$ per galaxy in these starburst galaxies with actively accreting SMBHs, which is on the order of $10^3$ times the overall observed TDE flare rate of approximately $10^{-5}{\rm yr}^{-1}$ per galaxy. Finally, while distinguishing TDEs in actively accreting galaxies may be challenging \citep{2020ApJ...903...17C, 2021MNRAS.507.6196I}, there has been interest in active galactic nucleus (AGN) variability, and in particular the possible role of ``changing-look" AGN as possible TDE candidates \citep{2021ApJ...907L..21D,  2023ApJ...959L..19D, 2024arXiv240612096W, WangLin2024}. 

Numerous mechanisms have been proposed to explain the statistics of TDE host galaxies. For example, arguments have been made that the rapid assembly of an accretion disk may be rare, and select more massive SMBHs or those with a pre-existing accretion flow \citep[e.g.][]{2015ApJ...809..166G,2019ApJ...881..113C}, with the implication that these might be preferentially found in the observed host galaxies. The distribution of stars around the SMBH has also been an area of close examination, including detailed analysis of particularly dense nuclear star clusters in post-starburst galaxies \citep[e.g][]{2016ApJ...825L..14S,2017ApJ...850...22L,2020MNRAS.497.2276P,2020ApJ...900...32P,2020A&ARv..28....4N}. Another possibility is the remaining presence of stellar disk structures, which are host to secular instabilities that drive stars to TDEs and interactions with the SMBH \citep{2009ApJ...697L..44M,2011ApJ...738...99M,2018ApJ...853..141M,2019ApJ...880...42W,2020ApJ...890..175F}. Finally, the possibility that TDE hosts are post-merger systems \citep{2021ApJ...908L..20H} has lead to arguments about the possible role of a secondary SMBH \citep[e.g.][]{2020ApJ...888L..14C}. 

In this paper, we consider the scenario of an accreting SMBH. Such a system is host to an accretion disk, and also ongoing tidal disruption events \citep{MacLeod2020}. If there is also an intermediate mass companion (IMC) object -- perhaps a less massive SMBH -- in the system, then 
its migration can dramatically modify the orbits of some stars \citep{yululin2007}.  A non-migrating IMC with an eccentric orbit can also
excite stars' eccentricity to parabolic and hyperbolic orbits through its sweeping secular resonance (SSR) 
\citep{Zheng2020, Zheng2021}. 
The mechanism of a sweeping secular resonance, driven by disk depletion, finds application across multiple astrophysical contexts. It has been used to model the dynamical evolution of planetesimals in the solar system and exoplanetary debris disks \citep{Nagasawa2003,Nagasawa2005,Zheng2017a, Zheng2017b} and has been extended to study the dynamical features of young stellar populations in the Galactic Center's central parsec \citep{Zheng2020,Zheng2021}. This includes explaining recent observations of the eccentricity–pericenter-distance distribution of S-stars (Zheng et al., in preparation). In these scenarios, an accreting disk plays a significant role in driving orbital precession for both the young stars and an IMC. A secular resonance occurs where the combined precession frequency for a star becomes comparable to the IMC's precession frequency. Since the disk's gravitational potential, and thus its contribution to precession, is determined by its surface density, the resonant location is inherently tied to the disk's instantaneous mass distribution. Consequently, as gas depletion reduces the disk's precessional influence, the secular resonance migrates, or sweeps, radially across a significant region.
This mechanism dramatically enhances stellar eccentricity excitation, driving stars toward parabolic orbits and subsequent tidal disruption.
We study the rate at which stars would be driven into TDEs by the combined presence of the IMC and the accretion flow. The coexistence of an IMC with the SMBH in such a system might have one of several origins. If the system were accreting in a post-merger phase, the IMC might represent a central SMBH of a less-massive galaxy \citep[e.g., as described by][]{2020ApJ...888L..14C}. Or, it is possible that an IMC black hole might naturally arise from runaway accretion onto stellar mass remnants in a sufficiently dense accretion flow \citep[e.g.][]{2012MNRAS.425..460M}.

The remainder of this paper is organized as follows. In Section \ref{sec:method}, we outline our modeling methodology and introduce an analytical framework for understanding how the SSR drives stars toward TDEs. We detail the components of the gravitational potential and the mechanisms of secular perturbation that govern the stellar orbital evolution in the vicinity of an SMBH and an IMC. In Section \ref{sec:sufficient}, we quantify the TDE enhancement rate attributable to the SSR mechanism across various galactic environments. We explore the parameter space defined by the IMC's mass ratio, orbital elements (eccentricity and semi-major axis), as well as the accretion disk's minimum mass and depletion timescale, to identify the conditions that maximize the SSR mechanism's efficacy. We further discuss the implications of these findings for the observed rates of TDEs in different types of galaxies. Finally, in Section \ref{sec:summary}, we summarize our conclusions and discuss the broader implications of the SSR mechanism for the study of TDEs and the dynamics of galactic nuclei. We highlight the key insights gained from our analysis and propose future research directions to further investigate the role of SSR in driving TDEs.

\section{Method}\label{sec:method}
We study the tidal disruption of stars induced by the SSR mechanism due to the depleting gas disk in  galactic nuclei that host a SMBH with an IMC. 

\begin{figure*}
\centering
\includegraphics[width=2\columnwidth]{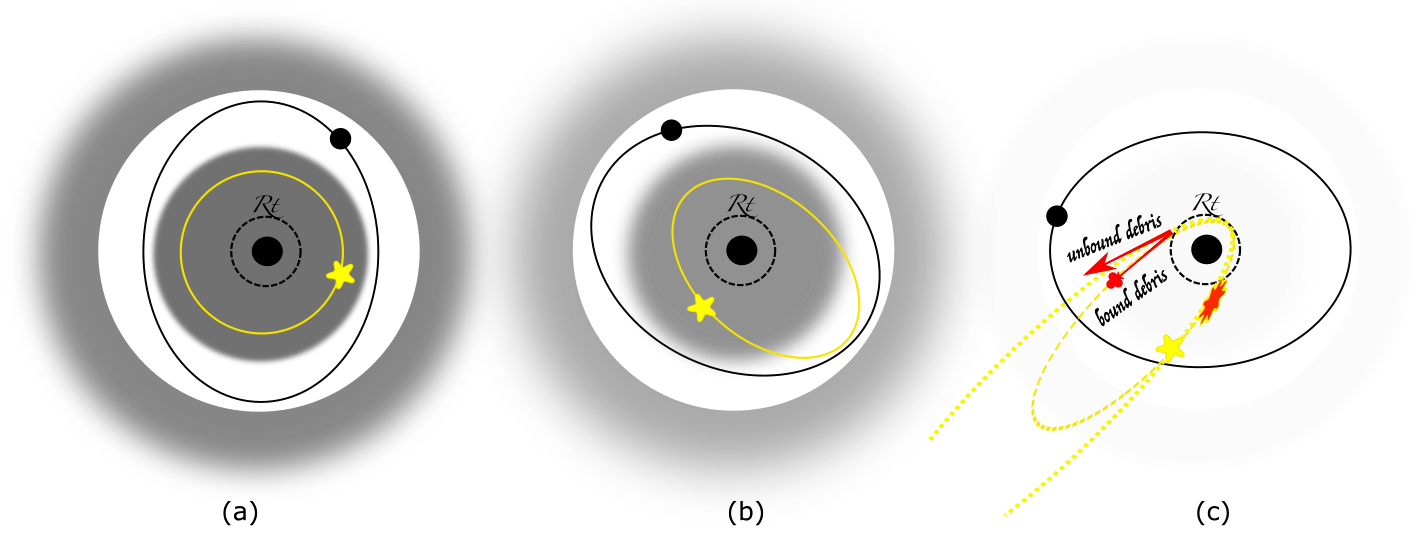}
\caption{The physical picture of producing TDEs under the sweeping secular resonance mechanism at each epoch. The black dots and yellow star refer to the central SMBH, the eccentric intermediate mass companion (hereafter IMC), and one representative disk star, respectively. The grey-shaded areas label a depleting gas disk. (a) The IMC opens a gap in the gas disk, and the representative disk star is initially in circular orbits and embedded in the gas disk. (b)The secular resonance of the IMC sweeps through an extensive region and excites the orbit of this disk star as the gas disk depletes over time. (c) If this highly excited representative disk star penetrates into the tidal disruption radii of the central SMBH (dashed circle), it will be shredded by the SMBH's strong gravity. Almost half of stellar debris remains gravitational bound and finally swallowed by the SMBH, which is associated with the observable TDEs. The rest of debris escapes from the dominated potential of the SMBH with an expanding tail.}
\label{fig:scenario}
\end{figure*}

\subsection{Gravitational potential}

The motion of the IMC and the stars are determined by a combination of potentials 
\begin{equation}
    \Phi=\Phi_{\bullet} + \Phi_{\rm IMC}+ \Phi_{\rm Bulge} + \Phi_{\rm Cluster}+ \Phi_{\rm Disk}
\label{eq:totalpot}
\end{equation}
where $\Phi_{\bullet}$ and $\Phi_{\rm IMC}$ are that due to the central SMBH and an satellite IMC respectively.  
Post-Newtonian gravity from a non-rotating host SMBH 
%, including (post)$^1$-Newtonian and (post)$^2$-Newtonian 
is taken into account in $\Phi_{\bullet}$ \citep{Kidder1995}.
We also consider the bulge potential of the galaxy. Generally, the bulges of disk galaxies can be approximated by the 
analytical model with a constant-density cores \citep{Tremaine1994}
\begin{equation}
    \Phi_{\rm Bulge} = - \frac{G M_{\rm Bulge} (R_{\rm Bulge} + 2 r)}{2 (R_{\rm Bulge} + r)^2} ,
\end{equation}
where $M_{\rm Bulge} = 10^{10} M_{\odot}$, $R_{\rm Bulge} = 0.6$ kpc are references to the mass and radii scaling factors for the bulge in our Galaxy \citep{Hernquist1990}.

The stellar cluster surrounding the central SMBH is also taken into account and can be simplified with a power-law spherical stellar 
number density ($\nu_\star$) and share a single stellar mass, $M_{\odot}$ as discussed as in the \cite{MacLeod2020}, 
therefore, the potential of such a stellar cluster can be described as
\begin{equation}
    \Phi_{\rm Cluster} = \frac{2 G M_{\bullet}}{(2-\gamma) r} \left(\frac{r}{R_{h}} \right)^{3 - \gamma} , 
    \label{eq:phicluster}
\end{equation}
where $\gamma = 1.5$ throughout this work, $R_{h} = G M_{\bullet} /  \sigma_{h}^2$ is a length scale describing the ``radius of influence"  of the central SMBH. The velocity dispersion is $\sigma_{h} = 2.3 (M_{\bullet}/M_{\odot})^{1/4.38}$ km/s, and we note that all settings are following the work of \cite{MacLeod2020}.

In addition, the gravitational potential $\Phi_{\rm d}$ of a hypothetical geometrically thin gaseous disk (in which the stars 
emerge) is determined by its poorly constrained surface density ($\Sigma$) distribution.  
We adopt the conventional $\alpha$ prescription for an effective viscosity $\nu$ \citep{shakura1973}
for a quasi steady-state disk such that the gas' radial velocity $U = -3 \nu/2 r = -3 \alpha h^2 \Omega r /2$
where $\Omega=\sqrt{G M_{\bullet} / r^3}$ is the Keplerian frequency, $h = c_s/ \Omega r$ is the aspect ratio,
and $c_s$ is the sound speed.  In a quasi steady state, the mass flux ${\dot M}$ is independent of $r$ although 
it can evolve with time such that 
\begin{equation}
    {\dot M}= 2 \pi \Sigma U r = 3 \pi \Sigma \alpha h^2 \Omega r^2 = \frac{3 \alpha h^3 M_{\bullet} \Omega }{Q}
\end{equation}
where $Q = h M_{\bullet} / \pi \Sigma r^2$ is the gravitational stability parameter.  In a self gravitating 
disk with ongoing star formation, $Q \simeq 1$ and $\alpha \simeq 1$ \citep{lin1988} such that $h \propto 
r^{1/4}$ and $\Sigma \propto r^{-3/2}$.

This radial distribution of $\Sigma$ is similar to that we have adopted in our previous models of protostellar 
disks \citep{Hayashi1985, Nagasawa2005, Zheng2017a, Zheng2020}, with a power-law distribution of the surface density as 
\begin{equation}
    \Sigma = \Sigma_0 \left(\frac{r}{R_0} \right)^{-3/2} e^{-t/\tau_{\rm dep}} ~\rm g/cm^2,
\label{eq:sigmadist}
\end{equation}
where $\tau_{\rm dep}$ is the depletion timescale of the gas nebula, $R_0 = 1000$ AU is a fiducial radius.
%, and $\Sigma_0 = 10000~{\rm g/cm}^2$ in the fiducial model.
After introducing a set of mass and length scaling by factors of $10^7$ and $10^3$ respectively, the problem can be rescaled between the galactic center and the Solar system contexts.

We assume the embedded stars do not have 
sufficient mass to open gaps near their orbit and their apsidal precession induced by the gas disk is dominated by the gas in 
their neighborhood.  The disk's gravitational force on them can be computed as 
\begin{equation}
f_{\rm emb, gas} = \frac{\partial \Phi_{\rm Disk} }{ \partial r} \approx - 4\pi G \Sigma ,
\label{eq:f_gas_star}
\end{equation}
which is similar to the settings in \cite{Ward1981, Nagasawa2005, Zheng2017a}.

%{\color {red}
Around AGNs where star formation is prevalent, there may also be a rich population of recently formed or captured 
and rejuvenated stars.  Their spatial density and mass distributions are determined by their formation, capture, accretion, and 
migration rates.  Although their profusion may affect TDE's occurrence rate, we assume that their contribution to the gravitational potential 
is negligible.
%}

Similar to the interaction between the proto giant planets and their natal protostellar disks,
the IMC can strongly perturb the structure of the disk near its orbit.  For models in which the IMC's orbital angular momentum vector
is parallel to that of the disk (co-orbiting models), we assume the presence of a gap around its orbit at
\begin{equation}
\begin{split}
a_{\mp} = a_{\rm IMC} (1 \mp e_{\rm IMC}) \left[ 1 \mp \left( {q \over 3} \right)^{1/3} \right] \ \ \ \  \\
{\rm where} \ \ \ \ q={M_{\rm IMC} \over M_{\bullet}} , \ \ \ \
\end{split}
\label{eq:ainout}
\end{equation}
$a_{\rm IMC}$ and $e_{\rm IMC}$ are the orbital parameters of the IMC.  The structure of this gap can modify the
disk's force on the IMC and precession rate of its orbit \citep{Ward1981, Nagasawa2005, Zheng2017a, Zheng2020} such that
\begin{equation}
\begin{split}
& f_{\rm gap, gas} =  \frac{\partial \Phi_{\rm Disk}^{\prime} }{ \partial r} \\
 & \approx  2\pi G \Sigma \sum_{l=0}^{\infty} 
%\left[ \frac{(2l)!}{2^{2l} (l!)^2} \right]^2 \left(\frac{2}{4l+1}\right)  \\
A_l B_l \left[ \frac{1}{2l+1} \left( \frac{r}{a_{+}}  \right)^{k}  - \frac{1}{2l} \left( \frac{a_{-}}{r} \right)^{k}  \right] ,
\end{split} 
\label{eq:f_gas_IMC}
\end{equation}
where $A_l = \left[ \frac{(2l)!}{2^{2l} (l!)^2} \right]^2$, $B_l = \frac{4l(2l+1)}{4l+1}$ and $k = \frac{4l+1}{2}$.

For the case where we consider the possibility that IMC's orbital angular momentum vector is anti-parallel to that of the disk
(counter-orbiting models), the IMC does not open up a gap in the gaseous disk, as opening a clear gap is significantly more difficult for a retrograde perturber \citep{Ivanov2015}.

\subsection{IMC's secular perturbation}
In Equation (\ref{eq:totalpot}), $\Phi_{\rm IMBH}$, $\Phi_{\rm Bulge}$, and $\Phi_{\rm cluster}$ are spherically symmetric and
$\Phi_{\rm Disk}$ is axisymmetric about the disk angular momentum axis.  With a finite eccentricity $e_{\rm IMC}$, IMC's 
perturbation $\Phi_{\rm IMC}$ introduces a torque which leads to angular momentum transfer between itself and the stars 
and modulation in their Runge-Lenz vector.  The magnitude of the stars' Runge-Lenz vector is their eccentricity and its
direction is their longitude of peri-astron.  Since this modulation is accumulative over many orbits, 
it is useful to adopt Laplace's approach in celestial mechanics by averaging periodic velocity and position changes over 
an orbital time scale. We consider the secular (long-term) evolution of the stars' orbital elements, including their semi 
major axis $a_\star$, eccentricity $e_\star$, inclination $i_\star$, longitudes of the peri-astron $\varpi_\star$, 
ascending node $\Omega_\star$.
%, mean longitudes $\lambda$ and mean anomaly $\vartheta$.  
This approach provides physical insights on the dominant contribution of various effects.  

We neglect the stars' feedback on the IMC's orbit such that its energy and angular momentum, semi major axis $a_{\rm IMC}$, 
and eccentricity $e_{\rm IMC}$ do not evolve with time.  Its longitude of peri-astron $\varpi_{\rm IMC}$ and ascending 
node $\Omega_{\rm IMC}$ precess at rates which evolve during the depletion of the disk.  To first order in $e_{\rm IMC}$, 
the rate of change in stars' eccentricity $e_\star$ and longitude of peri-astron $\varpi_\star$ under the IMC's secular 
perturbation can be approximated \citep{Murray1999, Nagasawa2003} by
\begin{align}
    \frac{d \xi}{d t} &= - \frac{q}{4} n_{\star}   \alpha \tilde{\alpha} b_{3/2}^{(2)} {\rm sin}~\eta, 
\label{eq:dxidt}
\end{align}
and 
\begin{align}
     \frac{d \eta}{d t} &= \left\{
     \begin{array}{ll}
       g_{\star} - g_{\rm IMC}  &   \text{co-orbiting case} \\
       g_{\star} + g_{\rm IMC}  &   \text{counter-orbiting case}
     \end{array} \right.
\label{eq:xieta}
\end{align}
where the eccentricity ratio is $\xi \equiv e_{\star}/e_{\rm IMC}$, the relative longitude of periapse is $\eta \equiv \varpi_{\star} - \varpi_{\rm IMC}$,  $g_\star$ and $g_{\rm IMC}$ are the stars' and IMC's precession frequencies respectively,  $b_{3/2}^{(2)}$ is the Laplace coefficient for the semi-major axis ratio, $\alpha \equiv {\rm min}(a_{\star}, a_{\rm IMC}) / {\rm max}(a_{\star}, a_{\rm IMC})$, $\tilde{\alpha} \equiv a_{\star} / {\rm max}(a_{\star}, a_{\rm IMC})$, and the orbital mean motion $n_{\star} = \sqrt{G M_{\bullet} / a_{\star}^3}$ in the limit $m_{\star} \ll M_{\bullet}$. 
Note that in the above expression
we consider two limiting cases in which the stars' orbit around SMBH in the same plane as the IMC, either in the same or opposite
directions.  

The precession rate $g_\star$ includes both an $\eta$-dependent (``Alternating Current", or AC) component and $\eta$-independent (``Direct Current", or DC) component.  
For most stars, the amplitude of the DC component is much larger than the AC component such that $g_{\star} - g_{\rm IMC}$ (or $g_{\star} + g_{\rm IMC}$) remains either positive or negative definite and their $\eta$ either 
liberates within some limited range or circulates between $0-2 \pi$. In this case, the $e_\star$ modulation is 
confined.  But at any given instance, the amplitude of AC component may be comparable or larger than that of the 
DC component such that it is possible for $g_{\star} - g_{\rm IMC}$ (or $g_{\star} + g_{\rm IMC}$) to vanish with
a non-zero $\eta$ and the $e\star$ (or equivalently $\xi$) to grow towards unity.  This state is normally referred 
to as secular resonance.  The location of the secular resonance is determined by the disk mass.  As the disk mass
decreases, it sweeps through a region away from the orbit of the IMC.  

%
%
%where $C = -b_{3/2}^{(2)}/b_{3/2}^{(1)}$.  
%
\begin{figure}
\centering
\includegraphics[width=1\columnwidth]{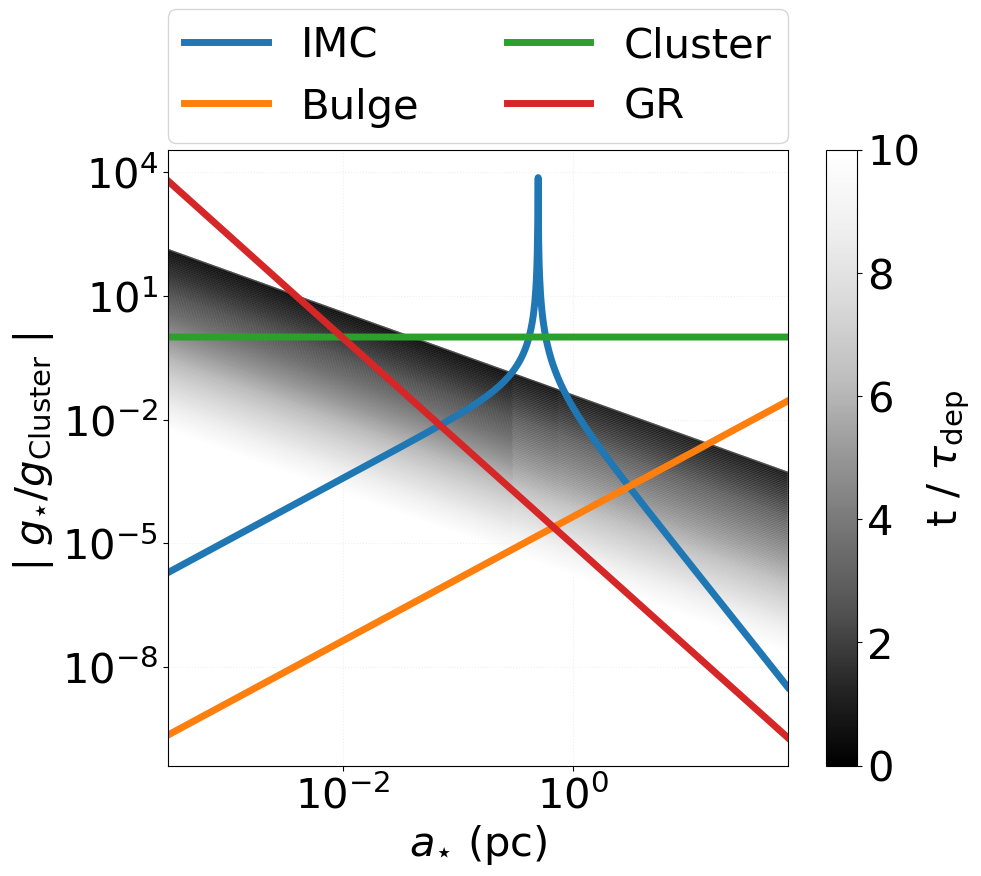}
\caption{
%{\bf Left panels:} 
The precession rates of the stars induced by different components of the gravitational potential, including that due to GR (from the central SMBH), stellar cluster, and the evolving gaseous disk at various depletion timescale (shown in grey color bar). 
These quantities are normalized by the cluster's induced precession rate. They are plotted as functions of the stars' initial 
location $a_{\star}$. 
%{\bf Right panels:} The time evolution of IMC's normalized precession rate, induced by GR, gaseous disk and stellar cluster. The normalization factor is the cluster's induced precession rate. The solid and dashed lines represent models with
%the IMC's orbital angular momentum vector parallel and anti parallel to that of the stars respectively.
The mass ratio between the SMBH and the IMC is set with $q = 10^{-2}$. The IMC is locating at $0.5$ pc with eccentricity 0.3. 
%The top panels and bottom panels are the results for initial and final stages, respectively. Blue shaded regions represent the sweeping %path of secular resonance of the IMC.
}
\label{fig:a_gcom_t}
\end{figure}

\subsection{Semi-analytic approximation of precession frequencies}

The above discussions clearly indicate the relevance of precession frequency for both the stars and IMC.  
As discussed as in \cite{Zheng2020}, the secular perturbation of an IMC (with semi-major axis $a_{\rm IMC}$) 
introduces a precession of the periastron of a disk star (with semi-major axis $a_{\star}$ and
mean motion $n_\star$) at a rate 
\begin{equation}
    g_{\star, \rm IMC} = \frac{n_{\star} q \alpha \tilde{\alpha}}{4} \left( b_{3/2}^{(1)} - \frac{e_{\rm IMC}}{e_{\star}}
    b_{3/2}^{(2)} {\rm cos}~\eta \right),
\end{equation}
where $b_{3/2}^{(1)}$ and $b_{3/2}^{(2)}$ are the Laplace coefficients for the semi-major axis ratio ($\alpha$).
Due to the large mass difference, the secular perturbation from the stars on the IMC can be ignored. 

To the lowest order of the eccentricity, the precession rate of the periastron included by the post-Newtonian gravity from the host SMBH can be simplified as
\begin{equation}
    g_{\rm GR} = \frac{3n}{2} \left(\frac{s}{a}\right) ,
\end{equation}
where $s = 2 G M_{\bullet} / c^2$ is the Schwarzchild radii of the central SMBH, and $c$ is the speed of light, and the orbital mean motion $n = \sqrt{G M_{\bullet} / a^3}$ as $m_{\star}/M_{\bullet} \ll q \ll 1$. We adopt $M_{\bullet} = 10^7 M_{\odot}$ and $s = 10^{-6}$ pc as the fiducial settings throughout the work.

As the location of central cusp stars satisfy $r \ll R_{\rm Bulge}$, the bulge potential can induce stars and IMC precess with an approximate rate
\begin{equation}
    g_{\rm Bulge} = - \frac{3n}{2} \left(\frac{M_{\rm Bulge}}{M_{\bullet}}\right) \left(\frac{a}{R_{\rm Bulge}}\right)^3  ,
\end{equation}
in the limit of small eccentricity.

The existence of the stellar cluster potential also provide additional perturbation both on the IMC and embedded stars, with the precession rate of the periastron's longitude as
\begin{equation}
   g_{\rm Cluster} = - 2 n \left(\frac{a}{R_h}\right)^{1.5}  = -2 {\sqrt{G M_{\bullet} \over R_h^3}} ,
\label{eq:gcluster}
\end{equation}
independent of $a$ in regions where 
$r \ll R_h$ (Eq. \ref{eq:phicluster}).  We adopt $R_h \approx 5.4$ pc in our fiducial model.

We define $M_{\rm Disk} = 4 \pi \Sigma_0 R_0^2 {\rm exp} (-t/\tau_{\rm dep})$ to be the characteristic gaseous disk mass at time $t$, the precession rate of the periastron's longitude induced by the gas disk can be classified as two types according to various situations, 
\begin{align}
   & g_{\rm Disk} \approx  -  \frac{n}{4} \left(\frac{M_{\rm Disk}}{M_{\bullet}}\right) \sqrt{\frac{a}{R_{0}}} Z_k, 
   \ \ \ \ \ \ \text{ where } \nonumber \\
   &  Z_k = \left\{
     \begin{array}{lr}
       1 &   \text{all stars \& an inclined IMC} \\
     % - \sum_{l=1}^{\infty} A_l B_l \left[  \left( \frac{a}{a_{+}}  \right)^k +  \left( \frac{a_{-}}{a} \right)^k  \right]  
     Z_k^{\prime} &   \text{ a co-orbiting IMC}    
      \label{eq:zk}
     \end{array} \right. 
\end{align}
$Z_k^{\prime} = - \sum_{l=1}^{\infty} A_l B_l \left[  \left( \frac{a}{a_{+}}  \right)^k +  \left( \frac{a_{-}}{a} \right)^k  \right]$, $a_-, a_+, k, A_l$, and $B_l$ are defined in Equations (\ref{eq:ainout}) \& (\ref{eq:f_gas_IMC}). 
%
%  $A_l = \left[ \frac{(2l)!}{2^{2l} (l!)^2} \right]^2$, $B_l = \frac{4l(2l+1)}{4l+1}$ and $k = \frac{4l+1}{2}$, which are referenced from the work of \cite{Ward1981, Nagasawa2005, Zheng2017a, Zheng2020}.

In comparison with the perturbation from the stellar cluster, the precession rate induced by various components is on the order of 
\begin{align}
   & \frac{\vert g_{\star,\rm IMC} \vert}{\vert g_{\star,\rm Cluster} \vert}  \simeq \mathcal O \left( q \frac{R_h^{3/2}}{a_{\star}^{3/2}} \right) ,  \nonumber \\
   &  \frac{\vert g_{\star, \rm Bulge} \vert}{\vert g_{\star,\rm Cluster} \vert}  \simeq \mathcal O \left(\frac{M_{\rm Bulge}}{M_{\bullet}} \frac{a_{\star}^{3/2}}{R_{\rm Bulge}^{3/2}} \frac{R_{h}^{3/2}}{R_{\rm Bulge}^{3/2}} \right) ,  \nonumber \\
   & \frac{\vert g_{\star, \rm GR} \vert}{\vert g_{\star,\rm Cluster} \vert}  \simeq \mathcal O \left(\frac{s^{5/2}}{a_{\star}^{5/2}} \frac{R_h^{3/2}}{s^{3/2}} \right) ,  \nonumber  \\
   &  \frac{\vert g_{\star, \rm Disk} \vert}{\vert g_{\star, \rm Cluster} \vert}
    \simeq \mathcal O \left( \frac{M_{\rm Disk}}{M_{\bullet}} \frac{R_0}{a_{\star}} \frac{R_h^{3/2}}{R_0^{3/2}} \right)  .
    \label{eq:g/gcluster}
\end{align}
These ratios indicate that, under the present settings, the stars' precession in the galactic nuclei are mainly affected by 
the surrounding stellar cluster potential, secular perturbation of the IMC, and the gaseous disk before it is greatly 
dissipated, while the influence of the bulge can be ignored as $a_{\star} \ll R_{h} \ll R_{\rm Bulge}$. The post-Newtonian contribution is only importantly at $a_\star \lesssim s^{2/5} R_h^{3/5}$, see also in the Figure \ref{fig:a_gcom_t}.

\begin{figure*}
\centering
\includegraphics[width=\columnwidth]{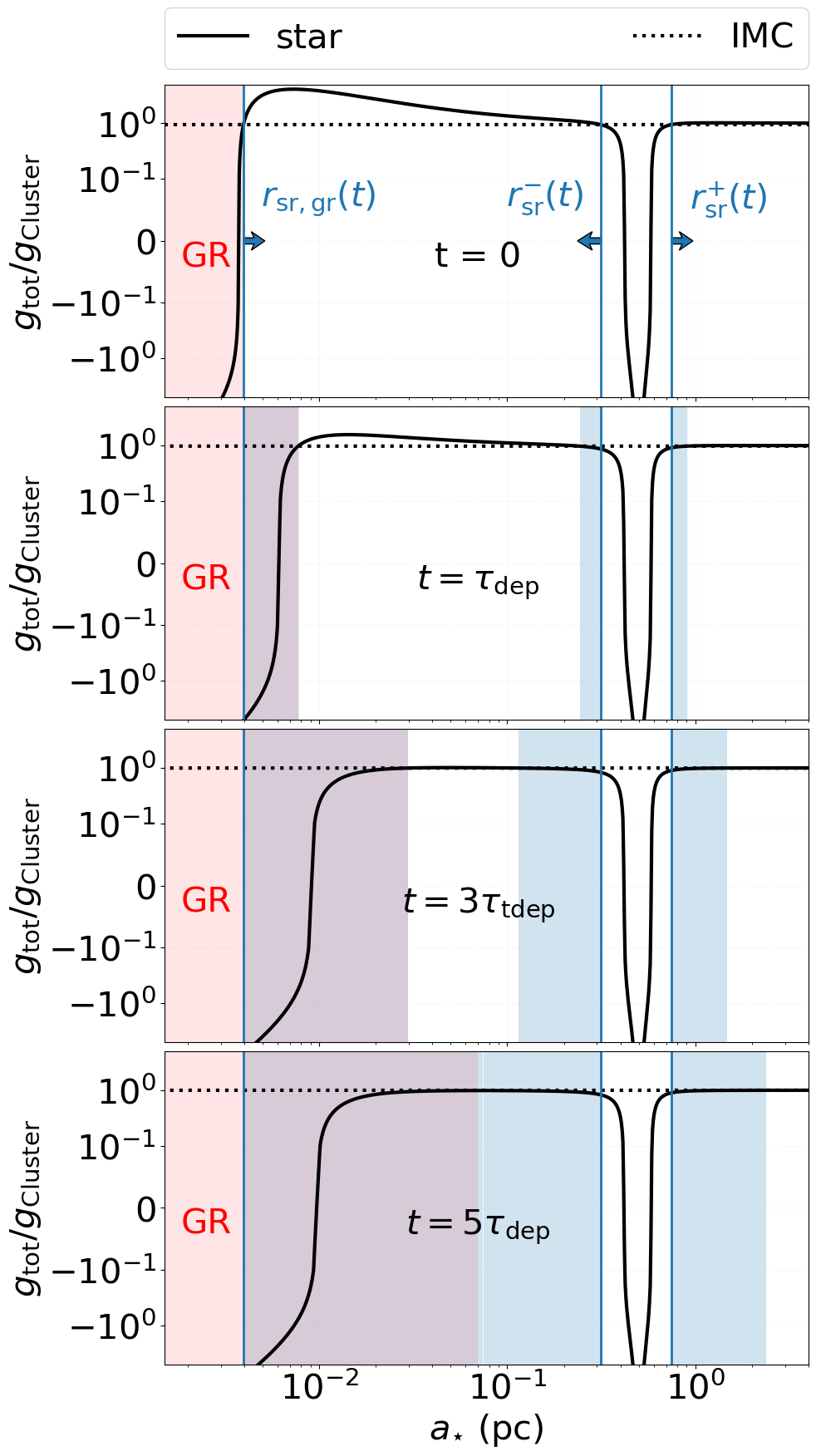}
\includegraphics[width=\columnwidth]{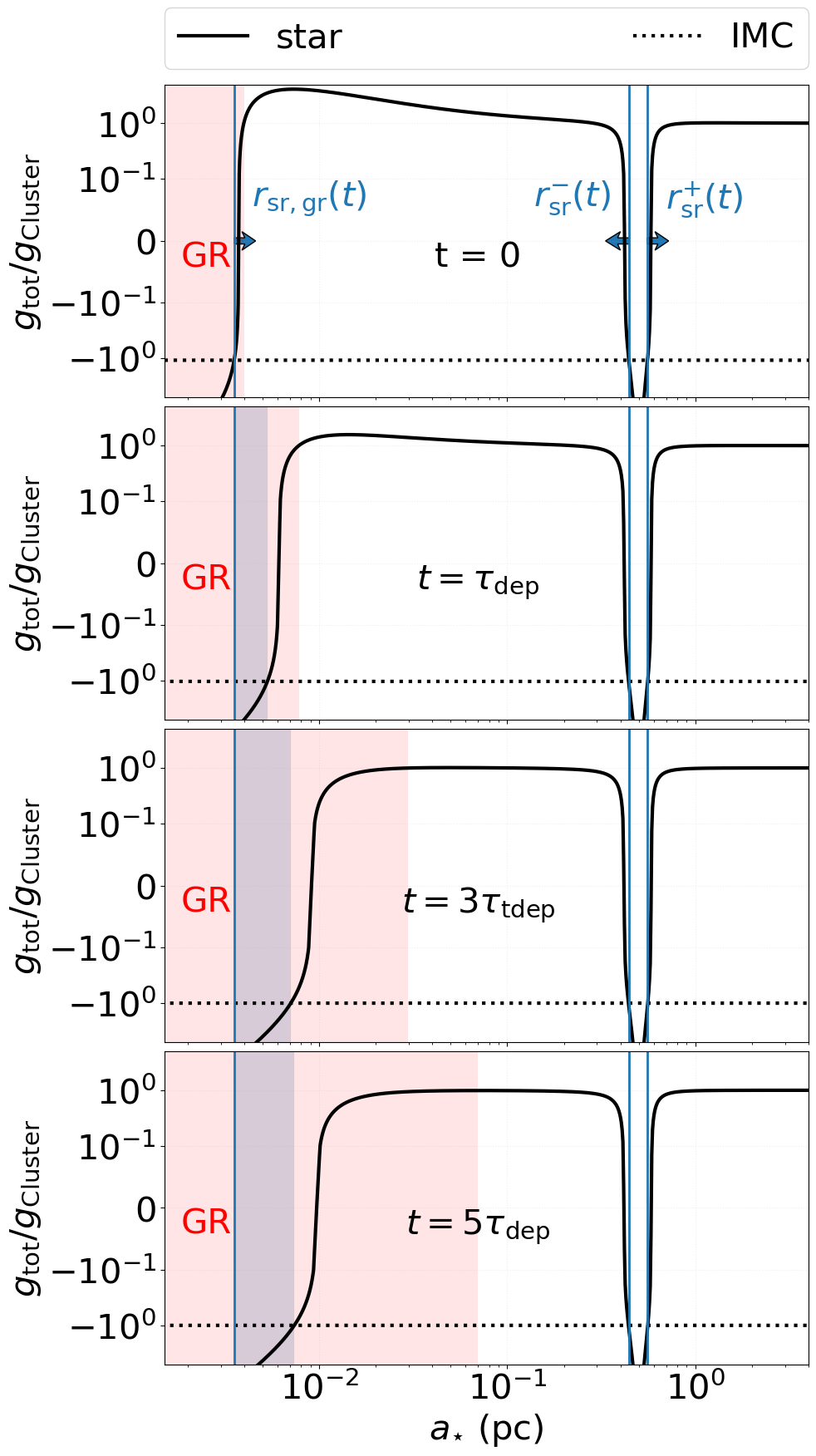}
\caption{The net precession rate ($g_{\rm tot}$) of the stars (solid black lines) and the IMC (dotted lines), normalized by the cluster-induced precession rate of the stars ($g_{\rm Cluster}$), as a function of the stars' initial semi-major axis $a_{\star}$. {\bf Left panels:} IMC's orbital angular momentum vector is parallel to that of the stars (clockwise co-orbiting, {\it CW} model); {\bf Right panels:} IMC's orbital angular momentum vector is 
antiparallel to that of stars (counter-clockwise, counter-rotating, {\it CCW} model). The mass ratio between the SMBH and the IMC is set with $q = 10^{-2}$. The IMC is locating 
at $0.5$ pc with eccentricity 0.3. Panels from top to bottom show successive stages of disk depletion.
The blue shaded regions denote the idealized sweeping path of IMC's secular resonance based on the criteria $g_{\star} \approx g_{\rm IMC}$ (Equations \ref{eq:gstar} and \ref{eq:gimc}). 
Arrows indicate how the resonance sweeps outward from the IMC's location, defined by $r_{\rm sr}^{-} (t)$ and $r_{\rm sr}^{+} (t)$.
Simultaneously, a resonance, $r_{\rm sr, gr} (t)$, sweeps outward from the GR-dominated region (red shaded, whose outer boundary is $a_{\star, \rm GR} = {\rm min}(a_{\star,\rm Disk}, a_{\star, \rm IMC})$, see also in Eqs. \ref{eq:as_disk} and \ref{eq:as_imc}) as GR precession begins to dominate over the diminishing gas disk precession.}
\label{fig:a_gtot}
\end{figure*}

Thus, the net precession rate of the star and IMC can be expressed as the sum of each component, following 
\begin{align}
    g_{\star} &= g_{\star, \rm IMC} + g_{\star, \rm Bulge} + g_{\star, \rm Cluster} + g_{\star, \rm GR} + g_{\star, \rm Disk}  \nonumber \\
              &= g_{\star, \rm DC} + g_{\star, \rm AC}  \xi^{-1} {\rm cos}~\eta,
\label{eq:gstar}
\end{align}
where
\begin{align}
    g_{\star, \rm DC} & \approx n_{\star} \left[ \frac{q}{4} \alpha \tilde{\alpha} b_{3/2}^{(1)} 
              + \frac{3}{2} \left(\frac{s}{a_{\star}}\right) - 2 \left(\frac{a_{\star}}{R_h}\right)^{1.5} - \frac{p}{4} \sqrt{\frac{a_{\star}}{R_0}}    \right],   \nonumber \\
    g_{\star, \rm AC} & \approx - n_{\star} \frac{q}{4}  \alpha \tilde{\alpha} b_{3/2}^{(2)}, \ \ \ \ \ \ {\rm and} \ \ \ \ \ \ 
    p = M_{\rm Disk} / M_{\bullet}
\label{eq:diskmassratio}
\end{align}
where $g_{\star, \rm DC}$ and $g_{\star, \rm AC}$ are the $\eta$-independent and $\eta$-dependent components 
of stars' precession frequency, and the disk-to-SMBH mass ratio $p$ exponentially decays over time.  For the IMC, 
\begin{align}
   & g_{\rm IMC} = g_{\rm IMC, \rm Bulge} + g_{\rm IMC, \rm Cluster} + g_{\rm IMC, \rm GR} + g_{\rm IMC, \rm Disk} \nonumber \\
                 & \approx n_{\rm IMC} \left[ \frac{3}{2} \left(\frac{s}{a_{\rm IMC}}\right) - 2 \left(\frac{a_{\rm IMC}}{R_h}\right)^{1.5} - \frac{p}{4} \sqrt{\frac{a_{\rm IMC}}{R_0}} Z_k  \right] ,
\label{eq:gimc}
\end{align}
respectively.  

We normalized the various components of $g_{\star, \rm DC}$ with respect to the contribution from the cluster. The radial distribution of these quantities is depicted in Figure \ref{fig:a_gcom_t}. To provide a basic understanding of each component, we consider a setting where the IMC is initially located at a semi-major axis $a_{\rm IMC} = 0.5$ pc and an eccentricity $e_{\rm IMC} = 0.3$. The mass ratio between IMC and the central SMBH is $q = 0.01$, the preliminary surface density is $\Sigma_0 = 12000~{\rm g/cm^2}$, and the initial mass ratio of the disk to the SMBH is $p = M_{\rm Disk} (t=0)/M_\bullet = 0.0017$. 
%{fig:a_t_precession}. 
The contributions from the IMC, GR, the bulge, and the cluster remain constant over time, whereas the contribution from the disk potential decreases as the disk depletes. 

%%%%%%%%%%%%%%%%%% added %%%%%%%%%%%%%%%%%%%%%%
To provide a clearer and more intuitive understanding of the dominant regions for each dynamical component, we have included Table \ref{tab:a_all}, which summarizes the definitions of key critical zones. The subsequent analysis of the secular resonance zone will be based primarily on a discussion of the competitive dominance between these components.

\begin{table*}[htbp]
\centering
\caption{Key symbols used in the secular resonance analysis, listed in order of appearance in the text.}
\label{tab:symbols}
\begin{tabular}{lll}
\hline
Symbol & Definition & Context \\
\hline
$a_{\star,\rm Disk}$ & Critical $a_{\star}$ where disk-induced precession comparable to GR-induced precession & $|g_{\star,\rm Disk}| = |g_{\star,\rm GR}|$ \\
\hline
$a_{\star,\rm IMC}$ & Critical $a_{\star}$ where IMC-induced precession comparable to GR-induced precession & $|g_{\star,\rm IMC}| = |g_{\star,\rm GR}|$ \\
\hline
$a_{\star,\rm GR}$ & Critical $a_{\star}$ where GR is dominated & min($a_{\star, \rm Disk}, a_{\star, \rm IMC}$) \\
\hline
$a^{\prime}_{\star,\rm Disk}$ & Critical $a_{\star}$ where a disk-induced-IMC precession comparable to IMC-induced-star precession & $|g_{\rm IMC, Disk}| = |g_{\star,\rm IMC}|$ \\
\hline
$a_{\star,\rm Cluster}$ & Critical $a_{\star}$ where Cluster potential is dominated in the {\it CCW} model &  $ \vert g_{\star,\rm IMC} \vert = 2 \vert g_{\rm Cluster}\vert$ \\
\hline
$a_{\star,\rm sr}$ & Critical $a_{\star}$ where secular resonant achieve &  $  g_{\star} = g_{\rm IMC}$ \\
\hline
$r^{-}_{\rm sr}(t)$ & Secular resonance sweeping location at $t$ epoch &  $  g_{\star} = g_{\rm IMC}$ \\
& inside the orbit of IMC &  $< a_{\rm IMC}$ \\
&  and far from the GR-dominated regime &  $>> a_{\star, \rm GR}$ \\
\hline
$r^{+}_{\rm sr}(t)$ & Secular resonance sweeping location at $t$ epoch &  $  g_{\star} = g_{\rm IMC}$ \\
&  outside the orbit of IMC &  $> a_{\rm IMC}$  \\
\hline
$r_{\rm sr, gr}(t)$ & Secular resonance sweeping location at $t$ epoch &  $  g_{\star} = g_{\rm IMC}$ \\
& around the GR-dominated regime &  $ \sim a_{\star, \rm GR}$  \\
\hline
$a^{-}_{\rm sr, f}$ & Final location of secular resonance sweeping  &  $  g_{\star} = g_{\rm IMC}$ \\
& inside the orbit of IMC   &  $ < a_{\rm IMC}$ \\
& that satisfy effective orbital excitation  & $\tau_{\rm para}/\Delta \tau_{\rm sr} \sim 1$\\
\hline
$a^{+}_{\rm sr, f}$ & Final location of secular resonance sweeping  &  $  g_{\star} = g_{\rm IMC}$ \\
& outside the orbit of IMC   &  $ > a_{\rm IMC}$ \\
& that satisfy effective orbital excitation  & $\tau_{\rm para}/\Delta \tau_{\rm sr} \sim 1$\\
\hline
\end{tabular}
\label{tab:a_all}
\end{table*}
%%%%%%%%%%%%%%%%%%%%%%%%%%%%%%%%%%%%%%%%%%%%

Figure \ref{fig:a_gtot} presents the net precession rates for a representative star ($g_\star$, solid lines) and the IMC ($g_{\rm IMC}$, dashed lines) at several evolutionary stages，$t = 0, 5 \tau_{\rm dep}, 7 \tau_{\rm dep}, 10 \tau_{\rm dep}$. The left panel shows the {\it CW} model, in which the orbital angular momentum vector of the IMC is aligned with that of the gaseous disk. The right panel shows the {\it CCW} model, where these vectors are anti-aligned.

Secular resonances occur when the differential precession rates between the stars and the IMC vanish.
The locations where the solid and dashed lines intersect in Figure \ref{fig:a_gtot}, indicate a secular resonance where $g_\star = g_{\rm IMC}$. These areas are marked by blue shaded regions. These regions, swept by the resonance over time, typically consist of three distinct zones, two of which are proximate to the IMC's location, and a third lies within the region where General Relativistic (GR) precession dominates.

A comparison of the panels of Figure \ref{fig:a_gtot} reveals that the secular resonance sweeps a broader radial range in the {\it CW} model than in the {\it CCW} model. In the anti-aligned {\it CCW} case, the precession rates are primarily governed by the stellar cluster potential (see also in Figure \ref{fig:a_gcom_t}) and remain largely unaffected by the sweeping secular resonance (SSR).

\subsection{Secular resonance}
Since $g_{\rm IMC}$ does not have any phase dependence, Equation (\ref{eq:xieta}) becomes
\begin{equation}
    {d \eta \over dt} =  \Delta g + g_{\star, \rm AC}  { {\rm cos}~\eta \over \xi} ,
    \ \ \ \ \ \ {\rm where} \ \ \ \ \ \ \Delta g= g_{\star, \rm DC} \mp g_{\rm IMC}
\label{eq:detadt}
\end{equation}
for a co-orbiting/counter-orbiting IMC.  In the limit that $\vert \Delta g \vert \equiv 
\vert g_{\star, \rm DC} \mp g_{\rm IMC}\vert
\gg \vert g_{\star, \rm AC} \vert$, $d \eta/dt$ vanishes only in the limit of very small $\xi$, i.e. 
$e_\star \ll 1$.  The solutions of Equations (\ref{eq:dxidt}) \& (\ref{eq:detadt}), $\xi$ and $\eta$,
undergo small amplitude liberation or circulation.  

There are special boundary conditions under which $\vert \Delta g \vert
\sim \vert g_{\star, \rm AC} \vert$ such that it is possible for $d \eta/dt$ to vanish while
$\eta$ keep to be constant and $d \xi /d t > 0$.
%$\eta \neq 1$ and $\xi \sim 1$. 
In this case, $\xi$ grows (Eq. \ref{eq:dxidt}) as $e_\star$ is 
excited towards unity.  This condition is referred to as {\it secular resonance} since the persistence
of some non-zero values of $\eta$ requires the IMC's precession rate to approximately match with that of stars 
under its secular perturbation as well as that due to other components of the gravitational potential.  As $\Phi_{\rm Disk}$ decreases with 
vanishing $g_{\rm Disk, \star}$ and $g_{\rm Disk, IMC}$ during the disk depletion, the location of 
secular resonances sweep away from the IMC (Fig. \ref{fig:a_gtot}).

\subsection{Necessary condition for secular resonance}
\label{sec:necessarysr}
With a substitution 
\begin{equation}
     \xi_{\rm max} \equiv {e_{\star, \rm max} \over e_{\rm IMC}} 
     \equiv \frac{q}{4} \frac{n_{\star}}{\vert \Delta g \vert} \alpha \tilde{\alpha} b_{3/2}^{(2)}   \ \ \ \ \ \ 
     {\rm and} \ \ \ \ \ \ \tau_{\rm ecc} \equiv \frac{1}{\vert \Delta g \vert},
\label{eq:emax}
     \end{equation}
Equations (\ref{eq:dxidt}) \& (\ref{eq:detadt}) reduce to 
\begin{equation}
    {\tau_{\rm ecc} \over \xi_{\rm max}} {d \xi \over d t} \simeq - {\rm sin} \  \eta \ \ \ \ \ \ {\rm and} \ \ \ \ \ \ 
    \tau_{\rm ecc} {d \eta \over d t} \simeq \frac{\Delta g}{\vert \Delta g \vert} - {\xi_{\rm max} \over \xi} {\rm cos} \ \eta.
\label{eq:dimensionless}
     \end{equation}
On the characteristic timescale $\tau_{\rm ecc}$, $\eta$ reaches a minimum or maximum with
$d \eta/ dt =0$ and
%$\xi \simeq \xi_{\rm max} {\rm cos} \ \eta$. 
$\xi \simeq \xi_{\rm max} \ {\rm sin} \ \eta$.
Thereafter $\eta$ and $\xi$ evolve together
either in the form of libration (with small-amplitude modulations in $\xi$ and $\eta$ 
about $\xi_{\rm max}$ and 0 respectively) or circulation (with $\eta$ ranging between 0 
and $2 \pi$).  In the limit  $\vert \Delta g \vert \geq \vert g_{\star, \rm AC} \vert$, 
$\xi_{\rm max}$ is relatively small such that $e_{\rm \star, max} \ll e_{\rm IMC}$.  
In contrast, in the proximity of the secular resonance where 
\begin{equation}  
\vert \Delta g \vert \leq \vert g_{\star, \rm AC} \vert, 
\label{eq:necessaryss}
\end{equation}
stars precess with the nearly same frequency as the IMC such they can attain relatively
large $\xi_{\star, \rm max}$ with the possibility of $e_{\star, \rm max} > e_{\rm IMC}$.
We refer Equation (\ref{eq:necessaryss}) as the {\it necessary condition for secular resonance}.

From Eqs. (\ref{eq:diskmassratio}), (\ref{eq:gimc}), \& (\ref{eq:detadt}), we find
\begin{equation}  
\Delta g \approx \Delta g_2-\Delta g_1; \\
%\ \ {\rm where} \ \ \ \ 
%\end{equation}
%\begin{equation}   
%\Delta g_1 = 
%{M_{\rm Disk} \over M_{\bullet}} \sqrt{\frac{GM_{\bullet}}{R_0^3}}  \left( {Z_k R_0\over 4 a_\star} \mp {Z_k^{\prime} R_0\over 4 a_{\rm IMC}} \right),
\Delta g_1 = n_0 \frac{p}{4} \left( {R_0\over a_\star} \mp {R_0\over a_{\rm IMC}} Z_k  \right),
\label{eq:ssr1}
\end{equation} 
where $n_0 = \sqrt{GM_{\bullet}/R_0^3}$.
\begin{align}      
& {\rm And} \ \ \ \ \ \ \ \ \ 
%Delta g_2= & \sqrt{G M_{\bullet} \over a_{\rm IMC}^3} \Bigg\{ {M_{\rm IMC} \alpha \tilde{\alpha} b_{3/2}^{(1)} \over 
%4 M_{\bullet}} \left( { a_{\rm IMC} \over a_\star} \right)^{3/2} \nonumber \\
%              & + \frac{3s}{2 a_{\rm IMC}} \left[ \left( {a_{\rm IMC} \over a_{\star}} \right)^{5/2} \mp 1 \right] 
%               - (2 \mp 2) \left( {a_{\rm IMC} \over R_h} \right)^{3/2}  \Bigg\} . 
\Delta g_2=  n_{\rm IMC} \Bigg\{ \frac{q}{4} \alpha \tilde{\alpha} b_{3/2}^{(1)} \left( { a_{\rm IMC} \over a_\star} \right)^{3/2} \nonumber \\
& + \frac{3}{2} \frac{s}{a_{\rm IMC}} \left[ \left( {a_{\rm IMC} \over a_{\star}} \right)^{5/2} \mp 1 \right] 
               - (2 \mp 2) \left( {a_{\rm IMC} \over R_h} \right)^{3/2}  \Bigg\} . 
\label{eq:ssr2}
\end{align}
Equations (\ref{eq:zk}) \& (\ref{eq:ssr1}) imply that $\Delta g_1$ is positive definite for both an co-orbiting and an
counter-orbiting (or inclined) IMC.  Secular resonance can only be established (with $\Delta g=0$) in regions where 
$\Delta g_2 \geq 0$.
The three components on the right hand side of $\Delta g_2$ (Eq. \ref{eq:ssr2}) are contributed by the IMC, GR, 
and the cluster potential through the dependence on $M_{\rm IMC}$, $s$, and $R_h$ respectively. 
For given values of $M_{\bullet}$, $M_{\rm IMC}$ (through $q$), $a_{\rm IMC}$ and $R_h$,  
the magnitude of $\Delta g_2$ is a function of $a_\star$.  Close to the IMC (at $a_\star \sim a_{\rm IMC}$),
$\Delta g_2$ is dominated by its secular perturbation. GR precession dominates at very small $a_\star$.

\begin{figure*}[ht!]
%\centering
\includegraphics[width=2\columnwidth]{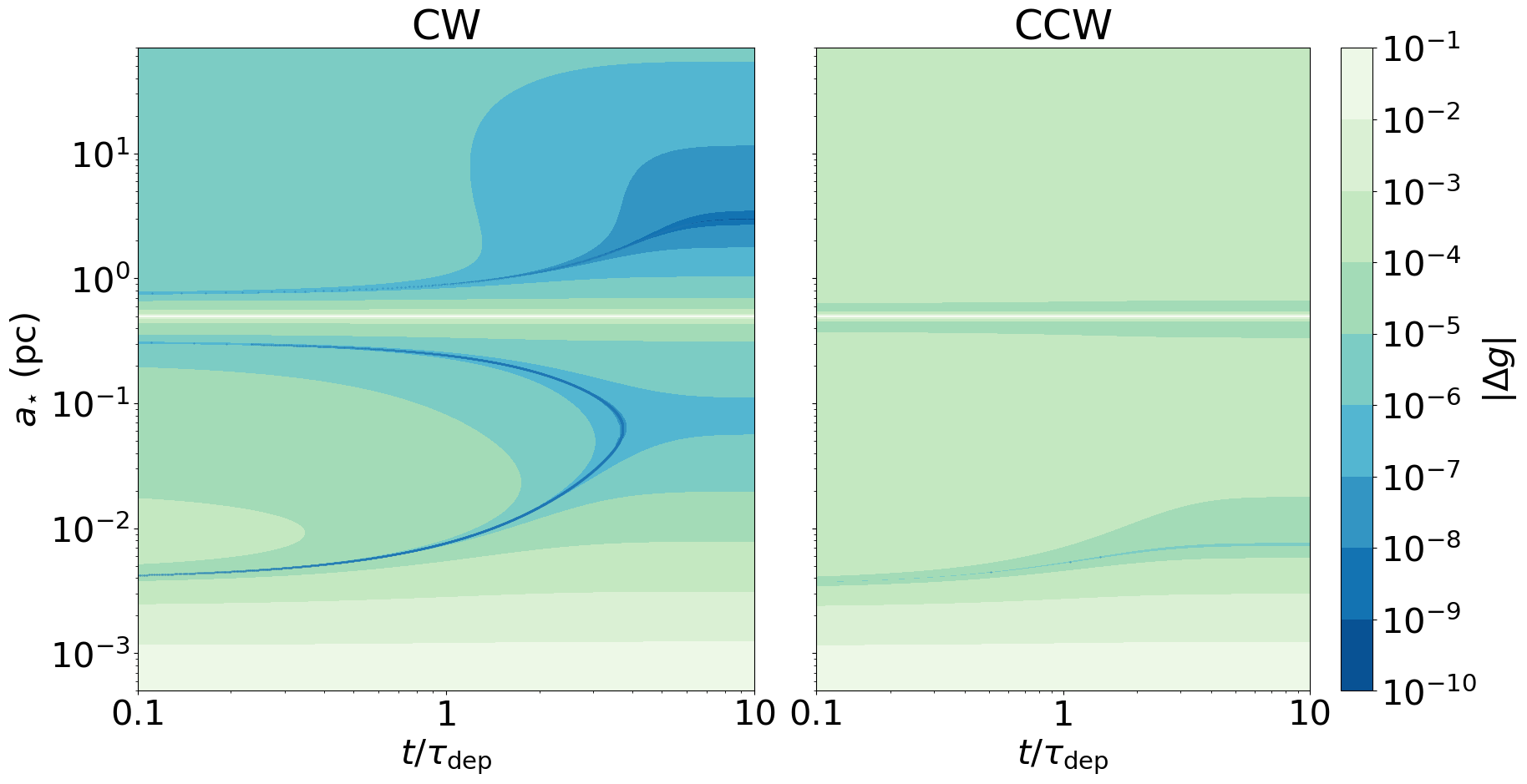}
\caption{The estimated secular resonance sweeping path according to the minimum value of $| \Delta g|$. Both the {\bf CW} (co-orbiting) and {\bf CCW} (counter-orbiting) cases are discussed in the left and right panels, respectively. 
}
\label{fig:ssr_deltg}
\end{figure*}

Figure \ref{fig:ssr_deltg} illustrates the approximate secular resonance sweeping paths as a function of time, derived from the calculation of $\Delta g$ in Equations \ref{eq:ssr1} and \ref{eq:ssr2}. The dark blue dots, marking the minimum values where $\Delta g \leq 10^{-9}$, represent these paths for an IMC with parameters $a_{\rm IMC} = 0.5$ pc, $e_{\rm IMC} = 0.3$, $q = 0.01$, and a disk profile defined by $p = 0.0017$.

Consistent with our previous analysis, we compare both the {\it CCW} and {\it CW} models. This figure clearly demonstrates why anti-aligned ({\it CCW}) orbits remain largely unaffected by the SSR, a result also seen in Figure \ref{fig:a_gtot}. In this configuration, the precession rate is dominated by the stellar cluster potential. The condition $\Delta g = \Delta g_2 - \Delta g_1 \sim 0$ is seldom met because the gravitational influence of the cluster ($\Delta g_2$) outweighs the dissipating effect of the gas disk ($\Delta g_1$).
Conversely, in the co-orbiting ({\it CW}) case, the cluster potential's contribution to $g_2$ is canceled, allowing the vanishing gas disk to play a decisive role. This leads to the adjustment $\Delta g_1 \approx \Delta g_2$, enabling the resonance condition.
As anticipated, the secular resonances originate within the gap region surrounding the IMC's orbit. Their inward propagation is halted near 0.1 pc due to the strong precession from GR effects.

%%%%%%%%%%%%%%%%%%%%%%%%%%%%%%%%%%%%%%%%%%%%%%%
\subsection{Extent of secular resonance zone}

Prior to the disk depletion, $M_{\rm Disk}$ and $\Delta g_1$ have maximum values. The GR precession
in $\Delta g_2$ is comparable to that due to the disk potential in $\Delta g_1$ at 
\begin{equation}
    {a_\star \over s} \sim {a_{\star, \rm Disk} \over s} = \left( {R_0 \over s} \right)^{1/3} \left( {6 \over p}\right)^{2/3}.
\label{eq:as_disk}
\end{equation}
The necessary condition for secular resonance (Eq. \ref{eq:necessaryss}) can generally be met
somewhere in the gas disk
%IMC 
dominated region (where  $a_\star \gtrsim  a_{\star, \rm Disk}$).  But, as 
$a_{\star, \rm Disk}$ increases with a diminishing
$M_{\rm Disk}$ during the disk depletion, the GR-dominated region expands.  Moreover, at $a_\star \leq a_{\star, \rm Disk}$,
the GR precession in $\Delta g_2$ exceeds the rate of stars' precession due to IMC's secular perturbation at 
\begin{equation}
    {a_\star \over s} \lesssim {a_{\star, \rm IMC} \over s} = {6 \over q}{1 \over  \alpha \tilde{\alpha} b_{3/2}^{(1)}} ,
\label{eq:as_imc}
\end{equation}
where the necessary condition for secular resonance (Eq. \ref{eq:necessaryss}) can no longer be satisfied ($ q \alpha \tilde{\alpha} b_{3/2}^{(1)} / 4 + 3 s / 2 a_{\star} < q \alpha \tilde{\alpha} b_{3/2}^{(2)} / 4 $).  Thus,  
secular resonance is excluded in the region interior to the minimum value of $a_{\star, \rm Disk}$ and $a_{\star, \rm IMC}$. In Figure \ref{fig:a_gtot}, the GR-dominated region is marked with a red shaded zone. As the gas disk disperses, the outer boundary of this region, $a_{\star, \rm GR} = {\rm min}(a_{\star, \rm Disk}, a_{\star, {\rm IMC}})$, progressively expands.

Since the magnitude of $g_{\rm Cluster}$ is constant (Eq. \ref{eq:gcluster}), it becomes
dominant cause for precession at large distance from the SMBH.  With the power-law 
$\nu_\star$ distribution we have used \citep{MacLeod2020} to calculation of $\Phi_{\rm Cluster}$
(Eq. \ref{eq:phicluster}),  the cluster's net contribution 
to the relative $\Delta g_2$ vanishes (Eq. \ref{eq:ssr2}) for a co-orbiting IMC. 
Since $\Delta g_2 < \Delta g_1$ for all
\begin{equation}
    {a_\star \over a_{\rm IMC}} \gtrsim {a_{\star, \rm Disk}^\prime \over a_{\rm IMC}} 
    = \left( {R_0 \over a_{\rm IMC}} \right)^{1/3} \left( \frac{q}{p} \right)^{2/3} \left( {\alpha \tilde{\alpha} b_{3/2}^{(1)}  
     \over \vert Z_k \vert}\right)^{2/3}
\end{equation}
where $Z_k < 0$ for an co-orbiting IMC, Eq. \ref{eq:zk}, secular resonance is excluded.  As the disk mass
decreases, the outer boundary $a_{\star, \rm Disk}^\prime $ of the secular resonance zone expands
and the near synchronization of the precession frequencies promotes peri-astron alignment and
the emergence of eccentric rings. 
%({\color{red} this is a process worth studying in the future.
%we may need to add some references here.}

For a counter-orbiting or an inclined IMC,  the cluster's potential introduces a net $\Delta g_2$
with a magnitude which is determined by $R_h$. Secular resonance would be excluded in the limit
that $\Delta g_2 \lesssim 0$ or at 
\begin{equation}
    {a_\star \over a_{\rm IMC}} \gtrsim \frac{a_{\star,{\rm Cluster}}}{a_{\rm IMC}} = 
    \left( \frac{q}{16} \alpha \tilde{\alpha} b_{3/2}^{(1)} \right)^{2/3} {R_h \over a_{\rm IMC} }
\end{equation}
regardless of the disk mass.  Consequently, the extent of secular-resonance zone 
is more confined for counter-orbiting than for co-orbiting IMCs.

\subsection{Necessary condition for parabolic-orbit excitation}

At the resonant location $a_\star = a_{\star, \rm sr}$, $\Delta g =0$.  
Although Eq. (\ref{eq:emax}) indicates $e_\star$ can attain arbitrarily large values,
the linear (low-$e_\star$) approximation in Eq. (\ref{eq:dimensionless}) breaks down.
Nevertheless, we can use these expressions as a guide for high-eccentricity excitation.

The stars' peri-center distance can only fall interior to the SMBH's tidal disruption 
radius when their $e_{\star, \rm max} \gtrsim 1$.  This requirement corresponds to 
$\xi_{\rm max} \gtrsim 1/e_{\rm IMC}$ (Eq. \ref{eq:emax}) or
\begin{equation}   
\vert \Delta g \vert \lesssim  \Delta g_{\rm crit} = \frac{q}{4} {n_{\star,\rm sr}} \alpha \tilde{\alpha} b_{3/2}^{(2)} e_{\rm IMC}
\label{eq:gcrit}
\end{equation}
in the proximity of $a_{\star, \rm sr}$, where $n_{\star,\rm sr} = \sqrt{G M_{\bullet}/a_{\star,\rm sr}^3}$.  This expression corresponds to the {\it necessary 
condition for parabolic-orbit excitation}.

In principle, all stars in the secular-resonant zone can be excited to attain $\vert \Delta 
g \vert \leq \vert \Delta g_{\rm crit} \vert$ (so that their $e_{\star, \rm max} > 
1 - R_{t}/a_{\star} \approx 1$ as $R_{t}/a_{\star} << 1$) are potential tidal disruption candidates. 
But, the characteristic time scale for their orbit to become parabolic (i.e. for their $e_\star$ to 
reach unity) is  
\begin{equation}
    \tau_{\rm para} \gtrsim 
%    \frac{\xi_{\rm ecc}} {e_{\star, \rm max}} = 
\frac{1}{\Delta g_{\rm crit}}
 %   \frac{1}{e_{\rm IMC}} \frac{1}{\vert g_{\star, \rm AC} \vert}
%    = \frac{4}{e_{\rm IMC} q n_{\star} \alpha \tilde{\alpha} b_{3/2}^{(2)}} 
= \left(\frac{q}{4} \alpha \tilde{\alpha} b_{3/2}^{(2)} \right)^{-1}  e_{\rm IMC}^{-1} n_{\star,\rm sr}^{-1} .
 %   = {4 M_{\bullet} \over e_{\rm IMC} M_{\rm IMC} 
 %   \alpha \tilde{\alpha} b_{3/2}^{(2)}} \sqrt {a_{\star, \rm sr}^3 \over G M_{\bullet}} .
    \label{eq:necessarypara}
\end{equation}
Provided the condition for the secular resonance (Eq. \ref{eq:necessaryss}) can be maintain over a few $\tau_{\rm para}$, 
 stars can attain nearly parabolic orbits leading to an enhanced rate of stars reaching the tidal disruption radius as an TDE (see further discussion below).
 
\subsection{Sufficient condition for parabolic-orbit excitation}
 
As $M_{\rm Disk}$ diminishes during disk depletion, the magnitude of $\Delta g_1$ 
decreases while the available time interval for the stars to be excited towards parabolic orbits 
\begin{align}
    \Delta \tau_{\rm sr} &\simeq { 2 \Delta g_{\rm crit} \over 
    \vert \partial \Delta g_1/\partial t \vert}  \nonumber \\ 
%    \simeq {\Delta g_{\rm crit} \tau_{\rm dep} \over \Delta g_1} 
    &= {q \over p} \left({R_0 \over a_{\star, \rm sr}}\right)^{3/2}
    { 2 \alpha \tilde{\alpha} b_{3/2}^{(2)} e_{\rm IMC} \over 
      (R_0/ a_{\star, \rm sr} \mp Z_k R_0/ a_{\rm IMC})} \tau_{\rm dep}
      \label{eq:tavailable}
\end{align}
increases.  
Parabolic-orbit excitation (Eq. \ref{eq:gcrit}) is efficiently accomplished in the limit $\tau_{\rm para} \leq \Delta \tau_{\rm sr}$.
From Equations (\ref{eq:necessarypara}) and (\ref{eq:tavailable}), we find
\begin{equation}
    {\tau_{\rm para} \over \Delta \tau_{\rm sr} }\simeq
    \frac{p}{q^2} {(R_0/ a_{\star, \rm sr} \mp Z_k R_0/ a_{\rm IMC}) \over (\alpha \tilde{\alpha} {b_{3/2}^{(2)}} e_{\rm IMC})^2} \left(\frac{a_{\star,\rm sr}}{R_0}\right)^3 \frac{1}{n_0 \tau_{\rm dep}} .
    %{M_{\rm Disk} M_{\bullet} \over M_{\rm IMC}^2 \tau_{\rm dep}} {\sqrt {R_0^3 \over GM_{\bullet}}} {a_{\star, \rm sr} ^3 \over R_0 ^3}
    %{(Z_k R_0/ a_{\star, \rm sr} \mp Z_k^{\prime} R_0/ a_{\rm IMC}) \over 
    %\alpha^2 \tilde{\alpha}^2 {b_{3/2}^{(2)}}^2 e_{\rm IMC}^2}.
    \label{eq:timeratio}
\end{equation}
This {\it sufficient condition for parabolic-orbit excitation} (i.e. relatively small 
$\tau_{\rm para} / \Delta \tau_{\rm sr}$) is more favorable in AGNs with 1) relatively 
massive IMC's, 2) relatively eccentric IMC's, 3) disks with relatively long depletion 
timescale, and 4) locations relatively close to the IMC.  

\subsection{Numerical verification}

To verify these analytic expressions, we carry out a series of numerical simulations with the open-source \textit{N}-body code REBOUND \citep{reinliu2012}, and choose 
the built-in MERCURIUS integrator \citep{rein2019}. The settings of model parameters are summarized in Table \ref{tab:parameters}, including: a) $M_{\bullet} = 10^7 M_\odot$, b) $M_{\rm IMC} = 10^{-2} M_{\bullet}$, 
c) $M_{\rm Disk} (0) = 1.7 \times 10^{-3} M_{\bullet}$, d) $R_0 = 1000$ AU, e) $R_{\rm h} = 5.4$ pc, f) $\tau_{\rm dep} = 10$ Myr, 
g) $a_{\rm IMC}=0.5$ pc, and h) $e_{\rm IMC} =0.3$. 
 We introduce a set of mass and length scaling by factors of $1 \times 10^7$ and $10^3$ respectively (see Table~\ref{tab:scale}) to highlight the similarity between the Galactic center and the Solar system. 

Based on the established differential impact of the secular resonance between models (see, e.g., Figures \ref{fig:a_gcom_t} and \ref{fig:ssr_deltg}), which demonstrates a significantly stronger effect in the aligned ({\it CW}) configuration compared to the anti-aligned ({\it CCW}) case, our simulation is initialized accordingly. 
At the onset of the simulation, we place a population of stars in circular orbits with a range of initial $a_\star$ in a disk which lies in the IMC's orbital plane and with parallel angular momentum vectors.
Guided by analytical estimates using the fiducial parameters in Table \ref{tab:parameters}, which indicate that the secular resonance can sweep through the entire target zone within approximately $5 \tau_{\rm dep}$ for the {\it CW} model, 
the stellar orbital evolution is computed over a duration of $5 \tau_{\rm dep}$.% with results represented in Figure (\ref{fig:a_tpara_sim}).

The simulations account for general relativistic effects through a post-Newtonian potential \cite{Kidder1995}. The inclusion of this GR potential is necessary to model the inner cutoff of the secular resonance, which ceases to propagate inward at the characteristic radius $a_{\star, \rm GR}$ where relativistic precession dominates.
To maintain focus on the core resonance mechanism, we omit the static potentials of the stellar cluster and the galactic bulge from our simulations. This simplification is analytically justified. The cluster potential induces nearly identical precessional rates on the IMC and the stars. Consequently, its net contribution to the secular resonance condition (Eq. \ref{eq:ssr2}) cancels in the {\it CW} model. Furthermore, the Bulge potential dominates only on significantly larger spatial scales (several parsecs) and thus has a negligible influence on the secular resonance sweeping within the sub-parsec region central to this investigation.

\begin{deluxetable}{cccc}
\tablenum{2}
\tablecaption{Scaling Galaxy to planetary system\label{tab:scale}}
\tablewidth{0.5pt}
\tablehead{
\nocolhead{} & \colhead{Observation} & \colhead{Scaling factors} & \colhead{Simulation}
}
%%\decimalcolnumbers
\startdata
%$\bf{Galactic Disk}$ \\
%$M_{\rm bulge}$ & $1 \times 10^{10} M_{\odot}$ & $4 \times 10^6$ & $2.5 \times 10^3~M_{\odot}$ \\
%$a_{\rm bulge}$ & 0.6 kpc & $1 \times 10^3$ & $1.2 \times 10^5$~AU \\
%$M_{\rm disk}$ & $4 \times 10^{10} M_{\odot}$ & $4 \times 10^6$ & $1 \times 10^4~M_{\odot}$ \\
%$a_{\rm disk}$ & 5 kpc & $1 \times 10^3$ & $1 \times 10^6$~AU \\
%$d_{\rm disk}$ & 0.3 kpc & $1 \times 10^3$ & $6 \times 10^4$~AU \\
%$M_{\rm halo}$ & $1 \times 10^{12} M_{\odot}$ & $4 \times 10^6$ & $2.5 \times 10^5~M_{\odot}$ \\
%$a_{\rm halo}$ & 20 kpc & $1 \times 10^3$ & $4 \times 10^6$~AU \\
%$\bf{Galactic Center super-massive Black Hole}$ \\
\\
$M_{\bullet}$ & $1 \times 10^7 M_{\odot}$ & $1 \times 10^7$  & 1 $M_{\odot}$ \\
$s$ & $1 \times 10^{-6}$~pc & $1 \times 10^3$  &  $2 \times 10^{-4}$~AU \\
$c$ & $ 3 \times 10^5~\rm km/s$  &  100 & $\sim 100~v_{\oplus}$ \\
%$\bf{Gas Disk}$ \\
\\
$\tau_{\rm dep}$ & $1$~Myr & 10  & $0.1$~Myr \\
$\Sigma_0$ & $10000 \rm~g/cm^2$ & 10  & $1000 \rm~g/cm^2$ \\
$R_0$ & 1000 AU & $1 \times 10^3$ & 1~AU \\
$H_0/R_0$ & 0.04 & 1 & 0.04 \\
%$\bf{Potential Intermediate mass Black Hole}$ \\
\\
$M_{\rm IMC}$ & $1 \times 10^5 M_{\odot}$ & $1 \times 10^7$ & $ 10 M_{\rm J}$ \\
%$s_{\rm IMC}$ & $3 \times 10^{4}$~km & $1 \times 10^3$  &  $30$~km \\
%$s_{\rm IMC}^\prime$ & 5000 AU & $1 \times 10^3$  &  $5$~AU \\
$a_{\rm IMC}$ & $0.5$~pc & $1 \times 10^3$ & $100$~AU \\
$e_{\rm IMC}$ & $0.3$ & 1 & $0.3$  \\
%$\bf{disk stars}$ \\
\\
$m_{\star}$ & $M_{\odot}$ & $1 \times 10^7$ &  $0.03 M_{\oplus}$ \\
%$s_{\star}$ & $1.5 \times 10^{6}$~km & $1 \times 10^3$  &  $1.5 \times 10^{3}$~km \\
$a_{\star}$ & $0.03-3$~pc & $1 \times 10^3$ & $6-600$~AU \\
%$v_{\star}$ & $\sim 200-600~\rm km/s$  &  100 & $\sim 0.1-0.3~v_{\oplus}$ \\
\enddata
\tablecomments{$M_{\bullet}$ %, M_{\rm bulge}$
 is the mass of SMBH, $c$ is the light velocity;
%, Galactic bulge; $a_{\rm bulge}$ is the characteristic length scales of the Galactic bulge; 
$M_{\rm IMC}, a_{\rm IMC}, e_{\rm IMC}$ are the mass, semi-major axis and eccentricity of the intermediate mass companion (IMC, which is either an IMBH or stellar cluster); $s$ %, s_{\rm IMC}$ 
 is the Schwarzchild radii of the SMBH; %and IMC; $s_{\rm IMC}^\prime$ is
%the softening parameter to represent IRS 13E as a stellar cluster, 
$\Sigma_0, H_0/R_0, \tau_{\rm dep}$ are the reference surface density, aspect ratio, characteristic depletion time of the gaseous disk, $m_\star, a_\star$ are the mass, initial semi-major axis of the stars.}
\tablecaption{}
\label{tab:parameters}
\end{deluxetable}
\begin{figure*}
\centering
\includegraphics[width=1.5\columnwidth]{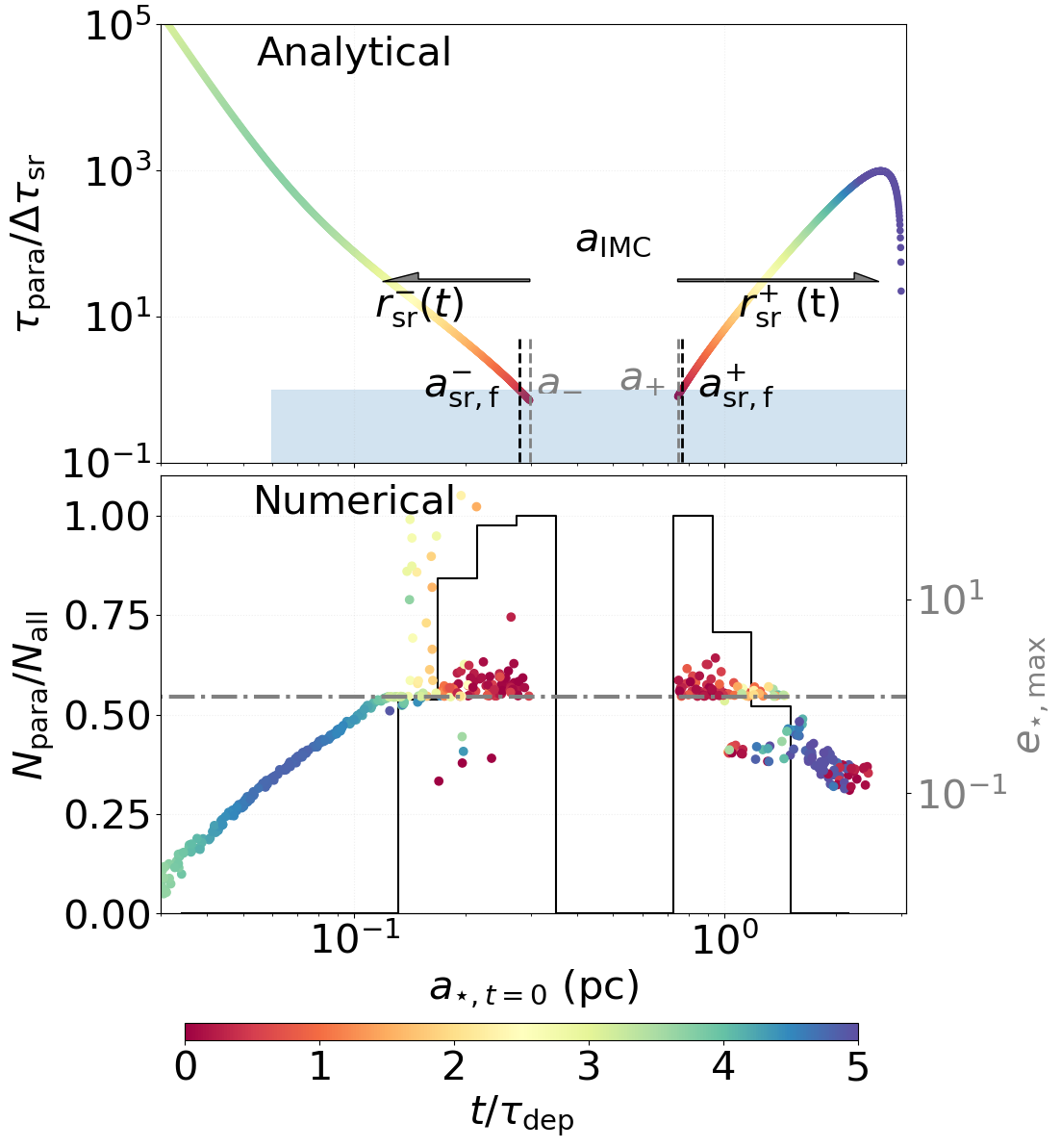}
\caption{{\bf Top:} The analytically estimated sufficient condition for parabolic-orbit excitation, $\tau_{\rm para} / \Delta \tau_{\rm sr} \leq 1$ (from Equation~\ref{eq:timeratio}), required for a star to achieve a maximum eccentricity $e_{\rm max} > 1$. {\bf Bottom:} Numerical results for the parabolic orbit occurrence frequency (left axis) and the maximum eccentricity (right axis), both as functions of the star's initial semi-major axis. The blue shaded region in the top panel indicates where the sufficient condition is satisfied. Colors represent time relative to the gas disk's depletion time. The system parameters are a mass ratio $M_{\rm IMC}/M_{\bullet} = 10^{-2}$, with the IMC located at 0.5 pc with an eccentricity of 0.1. We adopt a fixed gas disk depletion timescale $\tau_{\rm dep} = 10$~Myr.}
\label{fig:a_tpara_sim}
\end{figure*}
%

%{\color{blue} We may replace this paragraph with the two paragraphs below it.}  
Figure \ref{fig:a_tpara_sim} compares the analytical calculations based on Equation (\ref{eq:timeratio}) with the numerical simulations. It shows that in the parallel case, the secular resonance of one IMC with a certain location and eccentricity can sweep through a more extensive region (see also in Figure \ref{fig:a_gtot}). However, due to the constraints of $\tau_{\rm para} / \Delta \tau_{\rm sr}$, only a limited number of stars on a smaller scale can be highly excited to penetrate the tidal disruption region within certain periods.

The top panel of Figure \ref{fig:a_tpara_sim}  displays the ratio $t_{\rm para}/\Delta \tau_{\rm sr}$ (Eq. \ref{eq:timeratio}) as a function of the initial stellar semi-major axis, $a_{\star, t=0}$. The color mapping indicates the epoch $t_{\rm sr} ^\pm$ at which the secular resonance sweeps through a given orbital radius, $r_{\rm sr}^{\pm} (t)$. Consistent with the dynamics shown in Figures \ref{fig:a_gcom_t} and \ref{fig:ssr_deltg}, the inner resonance radius $r_{\rm sr}^{-}$ sweeps inward from $a_{-}$ staring at $t=0$. This inward propagation halts near 0.1 pc ($r_{\rm sr}^{\pm} (t)$) by $t \sim 4 \tau_{\rm dep}$ due to GR precession. This cessation creates an inflection point in the top panel, visible as a smooth transition in color around 0.1 pc. Within this region, the ratio $\tau_{\rm para}/\Delta \tau_{\rm sr}$ (Eq. \ref{eq:timeratio}), which gauges the efficiency of parabolic orbit excitation, increases monotonically as the stellar semi-major axis decreases.

Exterior to the IMC perturber, the outer resonance radius $r_{\rm sr}^{+}$ expands outward over time, as evidenced by the data point colors shifting smoothly from red ($t=0$) to blue ($t = 5 \tau_{\rm dep}$). However, the behavior of $\tau_{\rm para}/\Delta \tau_{\rm sr}$ diverges from this simple temporal trend. As Figure \ref{fig:ssr_deltg} illustrates, the outward propagation of the secular resonance stalls at approximately 3 pc. Consequently, the resonance dwell time, $\Delta \tau_{\rm sr}$, increases for stars at larger radii, causing $\tau_{\rm para}/\Delta \tau_{\rm sr}$ to peak and then decrease with increasing semi-major axis.

A key result is evident in the bottom panel of Figure \ref{fig:a_tpara_sim}. Although the fraction of stars excited to parabolic orbits, $N_{\rm para}/N_{\rm all}$, decreases with distance from the IMC perturber, this decline is not abrupt outside the domain defined by the {\it sufficient condition for parabolic-orbit excitation} ($\tau_{\rm para}/\Delta \tau_{\rm sr} \leq 1$), marked by $a_{\rm sr,f}^{-} \leq a_{\star} \leq a_{\rm sr,f}^{+}$.  This indicates that the adopted sufficient condition is highly restrictive, a significant population of stars with $\tau_{\rm para}/\Delta \tau_{\rm sr} \leq 10$ can still be perturbed onto parabolic orbits. We therefore conclude that our model, by relying on this strict criterion, provides a conservative lower bound on the total TDE rate enhancement driven by the SSR effect.

%\geq \frac{6 \left(s/a_{\rm I}\right)(\epsilon_{-}^{-1.5}-\epsilon_{-}) - q \epsilon_{-}^{1.5} \left[b_{3/2}^{(2)} (\epsilon_{-}) e_{\rm I}  - b_{3/2}^{(1)} (\epsilon_{-}) \right]}{(a_{\rm I}/R_{0})^{0.5} (Z_{k} - Z_{k}^{\prime} \epsilon_{-})} 

%%%%%%%%%%%%%%%%%%%%%%%%%%%%%%%%%%%%%%%%%%%%%%%%
\subsection{Sweeping of secular resonance}
\label{sec:ssr}

After the passage of secular resonance at $a_{\star, \rm sr}$, the necessary condition for parabolic excitation
can still be satisfied at a neighboring location with a smaller magnitude of $\Delta g_2$. This tendency for
relocation is referred to as {\it sweeping secular resonance}, and the direction of propagation is generally 
further away from the IMC.   In fact, the magnitude of $\tau_{\rm para} / \Delta \tau_{\rm sr}$ is small 
for relatively small $a_{\star,\rm sr}$ (Eq. \ref{eq:timeratio}). But, the secular resonance is excluded 
in the inner regions of the disk by GR precession (see \S\ref{sec:necessarysr}).

Both simulations and analytical models indicate that antiparallel alignment between the IMC's orbital angular momentum vector and that of the stars severely restricts the sweeping path. Therefore, to investigate the most efficient SSR mechanism for generating near-parabolic orbits, we focus exclusively on parallel orbital configurations in the following discussion.

The location of secular resonances ($r_{\rm sr}^\mp$ interior or exterior to $a_{\rm IMC}$) is determined by
the solution of $\Delta g=0$ (or $\Delta g_1 = \Delta g_2$ in Eq. \ref{eq:ssr1} \& \ref{eq:ssr2}).
Due to the stiff dependence of $g_{\star, \rm IMC}$ on distance from the IMC ($\vert a_\star - a_{\rm IMC} \vert$) and the competing contributions (from GR for small $a_\star$ and stellar cluster for large $a_\star$,  Fig. \ref{fig:a_gtot}). Taking the {\it sufficient condition for parabolic-orbit excitation} into consideration, the effective SR swept region is relatively close to the IMC.

The starting positions of the sweeping secular resonance at $t = 0$ are assumed to be the edges ($a_{\pm}$, Eq. \ref{eq:ainout})
%can be roughly estimated according to 
of the gas-free gap (the chaotic zone) around the orbit of the IMC.
However, it requires a minimum mass nebula disk ($p$, Eq. \ref{eq:diskmassratio}) for the co-orbiting case 
%({\color{red} we need to clarify this statement})
at $t = 0$ for satisfying $r_{\rm sr}^{\pm} = a_{\pm}$, following
%

 %   & p_{\rm min} \approx \left(p_1^N + p_2^N \right)^{1/N}  \ \ \  \ \ \ {\rm with} \nonumber \\
 % & p_1 = \frac{q \epsilon_{-}^{1.5} b_{3/2}^{(1)}(\epsilon_{-})}{ (Z_{k} - Z_{k}^{\prime} \epsilon_{-})} 
 %  \frac{R_{0}^{0.5}}{a_{\rm IMC}^{0.5}}, \ \ \ \     p_2  = \frac{q b_{3/2}^{(1)}(1/\epsilon_{+})}
 % {\epsilon_{+}^{2.5}  (Z_{k} - Z_{k}^{\prime} \epsilon_{+})} \frac{R_{0}^{0.5}}{a_{\rm IMC}^{0.5}},
%p_{\rm min} & \approx \left\{
%     \begin{array}{lr}
\begin{align}
& p_{\rm min}^{-} \approx
   q \frac{\epsilon_{-}^{1.5} b_{3/2}^{(1)}(\epsilon_{-})}{(1 - Z_{k}^{\prime} \epsilon_{-})} %+ 6 \frac{s}{a_{\rm IMC}} \frac{\epsilon_{-}^{-1.5} - \epsilon_{-}}{(1 - Z_{k}^{\prime} \epsilon_{-})}  
   \left(\frac{R_{0}}{a_{\rm IMC}}\right)^{0.5} ,  \nonumber \\ % &    \epsilon_{-} = r_{\rm sr}^{-} /a_{\rm IMC} = a_{-}/a_{\rm IMC} \\
      {\rm and} \quad \quad \quad
&  p_{\rm min}^{+} \approx  q \frac{b_{3/2}^{(1)}(1/\epsilon_{+})}{\epsilon_{+}^{1.5} (1 - Z_{k}^{\prime} \epsilon_{+})}
   %+ 6 \frac{s}{a_{\rm IMC}} \frac{\epsilon_{+}^{-1.5} - \epsilon_{+}}{(1 - Z_{k}^{\prime} \epsilon_{+})}  
   \left(\frac{R_{0}}{a_{\rm IMC}}\right)^{0.5} , 
\label{eq:pt0}
\end{align}
   
   %\\ %\nonumber%& \epsilon_{+} = r_{\rm sr}^{+}/a_{\rm IMC} = a_{+}/a_{\rm IMC}  \\
   % \end{array}
   %  \right,

where $\epsilon_{\mp} = r_{\rm sr}^{\mp}/a_{\rm IMC} = a_{\mp}/a_{\rm IMC}$. 
This minimum mass ratio, $p_{\rm min}$, is a simplified estimate derived from Equations~\ref{eq:ssr1} and~\ref{eq:ssr2} under the assumption that $a_{-} \gg a_{\rm sr, gr} \gg s$. As the IMC's secular resonance is unlikely to sweep through $a_{-}$ and $a_{+}$ simultaneously, maximizing the TDE rate requires the sweeping path to overlap both radii. We therefore define $p_{\rm min} = \max(p_{\rm min}^{-}, p_{\rm min}^{+})$. For subsequent calculations of the tidally disrupted star rate enhanced by sweeping secular resonance (SSR), we assume that all disks have a minimum mass $p_{\rm min}$ to guarantee that the resonance sweeps around the gap region. This allows us to derive the maximum possible TDE rate attributable to SSR.

\begin{figure*}
\centering
\gridline{\fig{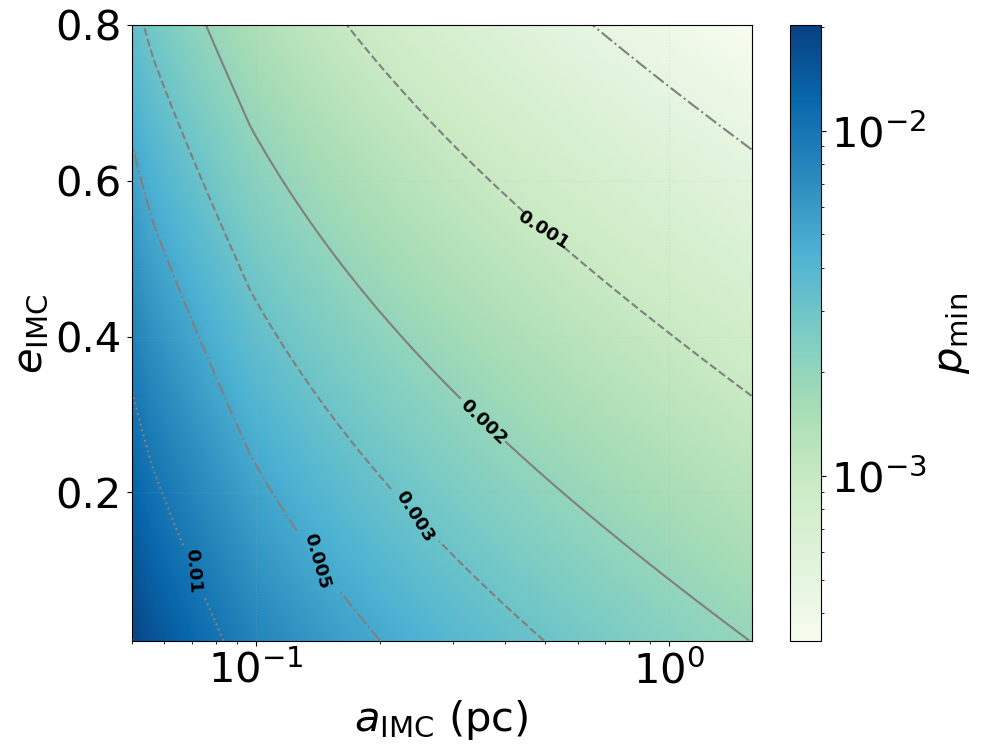}{0.5\textwidth}{(a) $q = 0.01$}
\fig{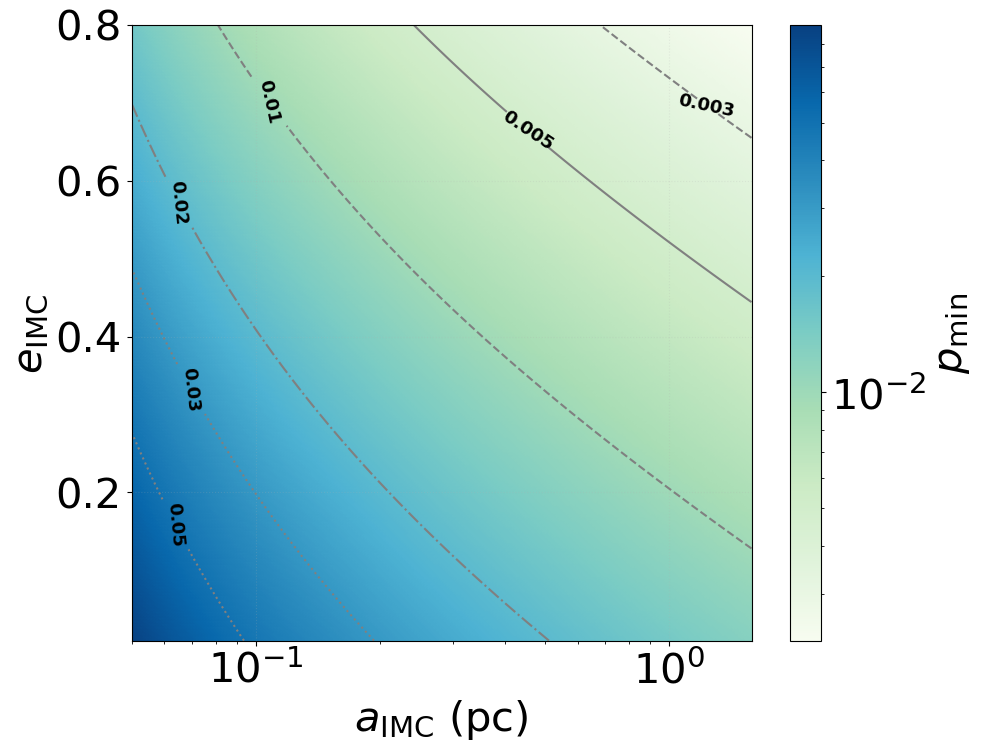}{0.5\textwidth}{(b) $q = 0.1$}}
\caption{The minimum initial nebular disk mass ratio ($p_{\rm min} = M_{\rm Disk} (t = 0)/ M_{\bullet}$) necessary to certify that secular resonance sweeping originates from the gap edges ($r_{\rm sr}^{\pm} (t = 0) = a_{\pm}$). 
This condition is required to induce strong secular resonance (SSR) and produce the maximum theoretical enhancement in the tidal disruption event (TDE) rate. The mass ratio of IMC ($q = M_{\rm IMC}/M_{\bullet}$) is set to be 0.01 in the {\bf Left} panel (a) and $0.1$ in the {\bf Right} panel (b). Grey lines mark the simplified estimates for $p_{\rm min}$ derived from Equation~(\ref{eq:pt0}).}
\label{fig:a_e_pmin}
\end{figure*}

Figure \ref{fig:a_e_pmin} shows the critical minimum initial gas disk mass ratio ($p_{\rm min} = M_{\rm Disk} (t=0)/M_\bullet$) required for the SSR mechanism to operate around the orbit of the IMC, $a_{\pm}$, as a function of the IMC's orbital parameters ($a_{\rm IMC}$, $e_{\rm IMC}$). The grey lines (solid, dashed, dotted, and dash-dotted) represent estimates derived from Equation \ref{eq:pt0}. The left (a) and right (b) panels correspond to mass ratios of $q = M_{\rm IMC}/M_{\bullet} = 0.01$ and 0.1, respectively. For the fiducial model with $q = 0.01$, $a_{\rm IMC} = 0.5$ pc, and $e_{\rm IMC} = 0.3$, a minimum disk mass ratio of $p_{\rm min} \approx 0.002$ is required to enable the maximum enhancement of the TDE rate.

Our analysis demonstrates that the minimum initial gas disk mass ratio, $p_{\rm min}$, is highly sensitive to the orbital parameters of the IMC and is typically smaller than the IMC mass ratio $q$. This result is physically reasonable, as can be seen from the brief discussion leading to Equation \ref{eq:g/gcluster}, where the ratio of the characteristic precession rates scales as $\vert g_{\star, \rm Disk} \vert / \vert g_{\star, \rm IMC} \vert \simeq \mathcal O \left( (p/q) (a_{\star}^{1/2} / R_0^{1/2}) \right)$. Consequently, in the regime where secular resonances occur—i.e., where $\vert g_{\star, \rm Disk} \vert \sim \vert g_{\star, \rm IMC} \vert$ and $a_{\star} \gg R_0$—the required initial mass of the gaseous disk is necessarily smaller than that of the IMC. An exception arises only when the radial location of the secular resonance is comparable to the fiducial radius $R_0$. As previously noted, achieving a maximally enhanced TDE rate requires that the secular resonance sweeps start from the gap region (where $a_{\pm} \gg R_0$) in the vicinity of the IMC. Accordingly, the minimum initial gas disk mass ratio we compute in Figure \ref{fig:a_e_pmin} is intimately linked to the orbital parameters of the IMC, specifically its semi-major axis $a_{\rm IMC}$ and eccentricity $e_{\rm IMC}$. A more distant IMC generally necessitates a substantially less massive gaseous disk, whereas an eccentric IMC tends to require a more massive disk. These trends are illustrated by the dashed, solid, and dotted contour lines in Figure \ref{fig:a_e_pmin}.

\section{Enhanced TDE occurrence rate}
%Sufficient condition for high-eccentricity excitation}
\label{sec:sufficient}

The effectiveness of sweeping secular resonance in high eccentricity excitation can be estimated from the requirement $\tau_{\rm para} 
/\Delta \tau_{\rm sr} \leq 1$ (Eq. \ref{eq:timeratio}). We introduce a probability function
${\mathcal {P}}(r_{\rm sr}^{\pm}(t))$ for parabolic-orbit excitation at $r_{\rm sr} ^\mp (t)$ as
secular resonances sweep through the nuclear stellar cluster such that
%is a factor to constraint the effective orbital excitation according to Equation (\ref{eq:t_ssr}), form as
%
\begin{align}
{\mathcal P}\left(r_{\rm sr}^{\pm}(t)\right) = \left\{
     \begin{array}{ll}
       0  &   \text{if} \quad \tau_{\rm para}^{\mp}(t) / \Delta \tau_{\rm sr} ^\mp > 1\\
    1  &   \text{if} \quad \tau_{\rm para}^{\mp}(t) / \Delta \tau_{\rm sr}
    ^\mp \leq 1
     \end{array}
     \right.
\end{align}
%({\color{red} note I have change the time criteria for $f$}).
%

\noindent
%{\color{red} we will comment out  two equations as in the origin version}
%For simplified, suppose $e_{\star, \rm max}$ can reaches unity within certain timescale, as
%%
%\begin{align}
%\tau_{\rm para}^{-} & = \frac{4}{q} \sqrt{\frac{(r_{\rm sr}^{-})^3}{G M_{\bullet}}} \left(\frac{a_{\rm IMC}}{r_{\rm sr}^{-}}\right)^2 %\frac{1}{ b_{3/2}^{(2)} (r_{\rm sr}^{-}/a_{\rm IMC})} \frac{1}{e_{\rm IMC}} \leq \Delta \tau_{\rm sr}^{-}  \nonumber \\
%\tau_{\rm para}^{+} & = \frac{4}{q} \sqrt{\frac{(r_{\rm sr}^{+})^3}{G M_{\bullet}}} \left(\frac{r_{\rm sr}^{+}}{a_{\rm IMC}}\right) %\frac{1}{ b_{3/2}^{(2)} (a_{\rm IMC}/r_{\rm sr}^{+})} \frac{1}{e_{\rm IMC}} \leq \Delta \tau_{\rm sr}^{+} .
%\label{eq:t_ssr}
%\end{align}
%%
We adopt a power-law stellar number density $\nu_{\star}$ which is self consistent \citep{MacLeod2020} with 
$\Phi_{\rm Cluster}$ (Eq. \ref{eq:phicluster}).
As the secular resonances sweep through the cluster during
depletion of the gaseous disk, the number frequency of tidal disruption candidates (those with $e_\star 
\rightarrow 1$) due to the sweeping secular resonance can be estimated as
\begin{align}
   & {\dot N}_{\rm ssr, TDE} (t)  =  \nonumber \\  
    & \frac{{\mathcal P}\left(r_{\rm sr}^{-}(t)\right)} {\delta t}
    \int_{-\pi/2}^{\pi/2}  \left[\int_{r_{\rm sr}^{-}(t+\delta t)}^{r_{\rm sr}^{-}(t)} \nu_{\star} 
    2 \pi r^2 \mathrm{d} r \right]  \mathrm{sin}~i~H(i) \mathrm{d} i \nonumber \\
    & + \frac{{\mathcal P}\left(r_{\rm sr}^{+}(t)\right)}{\delta t} \int_{-\pi/2}^{\pi/2} 
    \left[\int_{r_{\rm sr}^{+}(t)}^{r_{\rm sr}^{+}(t+\delta t)} \nu_{\star} 2 \pi r^2  \mathrm{d} r \right] 
    \mathrm{sin}~i~H(i) \mathrm{d} i ,  
\end{align}
%
%{\color{red} I changed $\nu_{\rm ssr, t}$ to ${\dot N}_{\rm ssr, TDE}$.  Also $h$ is used to replace $g$ because $g$ is used for precession frequency.}
%where the stellar number density is arranged according to a power-law distribution in radius and in the form as in \cite{MacLeod2020, 2013CQGra..30x4005M}
%\begin{equation}
%\nu_{\star} = \frac{3}{4 \pi} \frac{M_{\bullet}}{M_{\star}} %\left(\frac{r}{R_{h}}\right)^{1.5} r^{-3} ,
%\end{equation}
where $H(i)$ describes the mutual inclination $i$ (between the orbits of stars and the IMC) dependent excitation. 
The mutual inclination $i$ can, in principle, span from $-\pi/2$ to $\pi/2$. However, our analysis focuses on the sub-population  which are most susceptible to the sweeping secular resonance, these stars are likely initially embedded within the gaseous disk. The gas disk has a scale height that follows the relation $h = h_0(r/R_0)^{1/4}$. Consequently, we assume 
\begin{align}
H(i) = \left\{
     \begin{array}{ll}
       0  &   \text{if} \quad  {\rm sin}~i > h~~~{\rm or}~~~ {\rm sin}~i < -h \\
    1  &   \text{if} \quad -h \leq {\rm sin}~i \leq h 
     \end{array}
     \right.
\label{eq:hi}
\end{align}

There is no simple analytic approximation for
the effective secular resonance region for inclined cases, making this regime hard to estimate, not to mention 
the Kozai effect for $i > 39^{\circ}$ circumstances. 
%({\color{red} we should explain why it is ignorable.} 
%Fortunately, in this study, our primary focus is on the enhanced TDE rate resulting from the sweeping secular resonance.
Consequently, modeling the initial stellar distribution solely within the gas-aligned zone (in Equation \ref{eq:hi}, where the secular resonance operates, following our analytical estimation ) yields a conservative, lower-limit estimate of the TDE enhancement rate. The actual TDE rate driven by the sweeping secular resonance may therefore be greater.
%Fortunately, the numerical tests in the Figures \ref{fig:a0_rmin_inc} and \ref{fig:a0_rmin_n} 
%indicate that $h(i)$ can be around unity.
Finally, $r_{\rm sr} ^\mp (t)$ is the resonant location at time $t$ (see Figure \ref{fig:ssr_deltg}), $\delta t$ is a time interval.

Based on these approximations, the total enhanced tidal disruption rate during the SSR can be simplified as 
\begin{align}
%\nu_{\rm ssr, t} 
& {\dot N}_{\rm ssr, TDE}  \approx  \nonumber \\
& 6 \times 10^{-4} \left(\frac{M_{\bullet}}{M_{\star}}\right) \Bigg\{\frac{a_{-}^2}{R_h^{3/2} R_0^{1/2}} \left[ 1 - \left(\frac{a_{\rm sr,f}^{-}}{a_{-}}\right)^2 \right] \frac{1}{\Delta t_{-}}  \nonumber \\
& + \frac{a_{+}^2}{R_h^{3/2} R_0^{1/2}} \left[\left(\frac{a_{\rm sr,f}^{+}}{a_{+}}\right)^2 -1 \right] \frac{1}{\Delta t_{+}} \Bigg\} ,
%& + \left[\left(\frac{r_{\rm sr}^{+}(t+\delta t)}{R_h}\right)^{3/2} \left(\frac{r_{\rm sr}^{+}(t+\delta t)}{R_0}\right)^{1/2}  \nonumber \\
%& - \left(\frac{r_{\rm sr}^{+}(t)}{R_h}\right)^{3/2} \left(\frac{r_{\rm sr}^{+}(t)}{R_0}\right)^{1/2}\right] {\mathcal P}\left(r_{\rm sr}^{+}(t)\right) \Bigg\} , \\
\label{eq:v_ssr}
\end{align}
%& = \frac{2 M_{\bullet}}{M_{\star} \delta t} \left[\left(\frac{a_{-}}{R_h}\right)^{3/2} - \left(\frac{a_{\star, \rm sr}^{-}}{R_h}\right)^{3/2} \nonumber \\
%& + \left(\frac{a_{\star, \rm sr}^{+}}{R_h}\right)^{3/2} - \left(\frac{a_{+}}{R_h}\right)^{3/2} \right] 
where $a_{\rm sr,f}^{\mp}$ denotes the final locations of the secular resonance sweeping inside and outside the orbit of the IMC that satisfy the effective orbital excitation, fulfilling the {\it sufficient condition for parabolic-orbit excitation} (see also Equation \ref{eq:necessarypara}). The corresponding time intervals from $a_{\mp}$ to $a_{\rm sr,f}^{\mp}$ are denoted by $\Delta t_{\mp}$.

%Each panel shows the parameter plane of $a_{\rm I}$ and $e_{\rm I}$, $q$ is varied in different columns of panels range from $10^{-4}$ to $10^{-2}$, and $\tau_{\rm dep}$ is varied in different rows range from $0.1$ Myr to $10$ Myr. 

\begin{figure*}
\centering
\includegraphics[width=2\columnwidth]{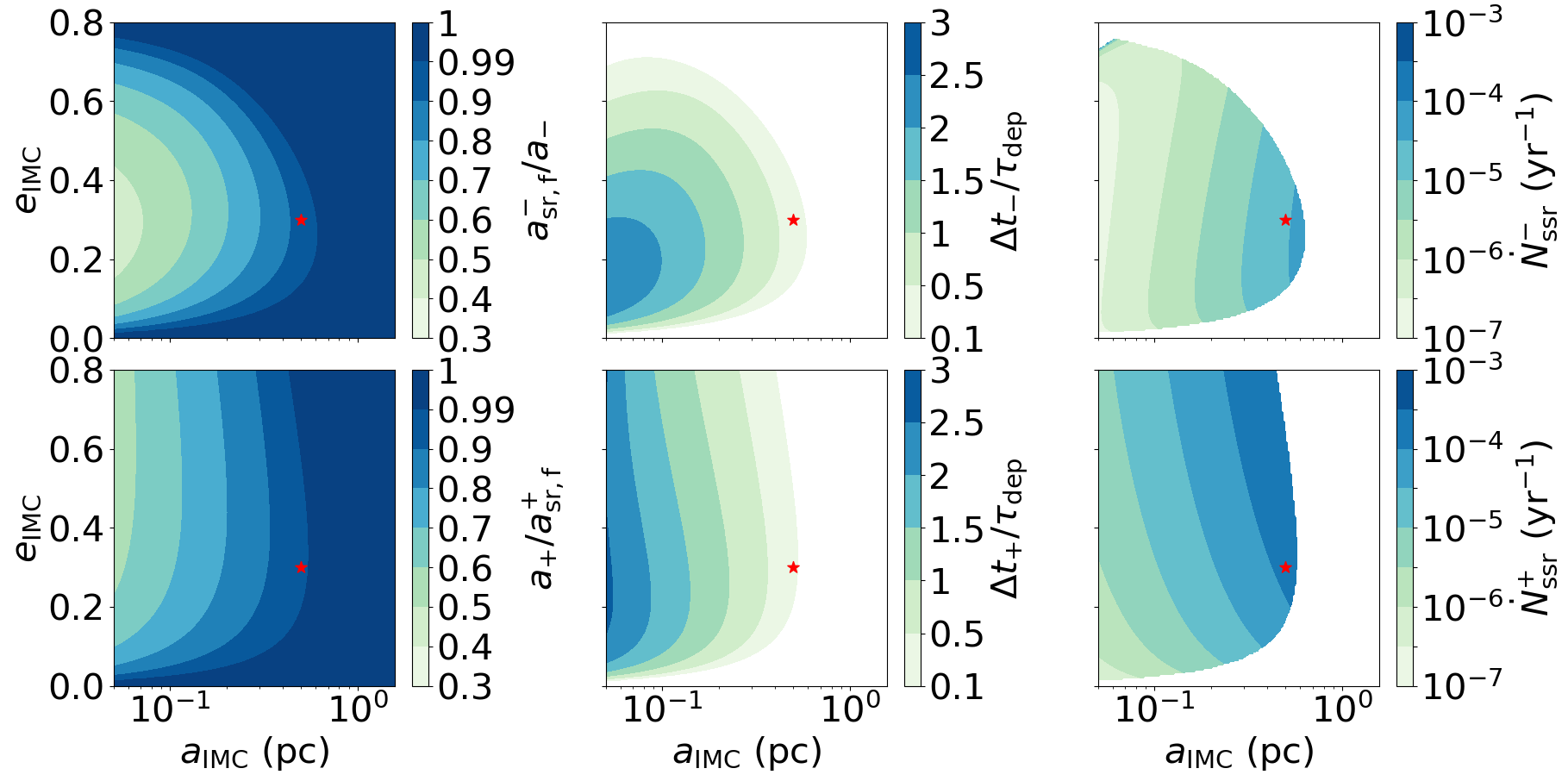}
\caption{TDE rate enhancement via the SSR mechanism driven by an IMC. Panels display the effective radii and active time intervals of the SSR regions, along with the corresponding TDE rate increase, both inside (top, labeled $-$) and outside (bottom, labeled $+$) the IMC's orbit, as defined in Equation~(\ref{eq:v_ssr}). Results are shown as a function of the IMC's semi-major axis $a_{\rm IMC}$ and eccentricity $e_{\rm IMC}$, with parameter variations indicated by the color bars. The fiducial model $(a_{\rm IMC} = 0.5, {\rm pc}, e_{\rm IMC} = 0.3)$ is highlighted with a red star. The model is computed under the following fixed parameters with IMC mass ratio $q = 10^{-2}$, gas disk depletion timescale $\tau_{\rm dep} = 10$ Myr, initial disk mass ratio $p = M_{\rm Disk}(t=0)/M_{\bullet}$ from Equation~(\ref{eq:pt0}).}
\label{fig:a_e_q0.01_sr}
\end{figure*}

In Figure \ref{fig:a_e_q0.01_sr}, we illustrate the dependence of the effective SSR regions ($a_{\rm sr, f}^{\mp}$) and the effective time span ($\Delta t_{\pm}$) on the orbit of the IMC, as described by Equation \eqref{eq:v_ssr}. These parameters are crucial for estimating the TDE enhancement rate due to the SSR mechanism. For this illustration, we adopt $q = 0.01$ and $\tau_{\rm dep} = 10$ Myr as example values.

The top panels of Figure \ref{fig:a_e_q0.01_sr} depict the inward sweep of the sufficient SR excitation region, ranging from $a_{-}$ to $a_{\rm sr, f}^{-}$, along with the associated time span $\Delta t_{-}$ and the induced TDE rate in this region. These are presented in three separate panels, each labeled with distinct color bars representing $a_{\rm sr, f}^{-}/a_{-}$, $\Delta t_{-}$, and $\dot{N}_{\rm ssr}^{-}$, respectively. The bottom panels, in contrast, focus on the region outside the orbit of the IMC. Here, three panels display the dependence of $a_{+}/a_{\rm sr, f}^{+}$, $\Delta t_{+}$, and $\dot{N}_{\rm ssr}^{+}$ on the IMC's orbital parameters.

Our analysis reveals that the effective SSR region is highly sensitive to the orbital parameters of the IMC. Specifically, the extent to which the IMC's secular resonance can sweep inward relative to the inner gap region, as indicated by $a_{\rm sr, f}^{-}/a_{-}$, is influenced by both the IMC's orbital eccentricity ($e_{\rm IMC}$) and its semi-major axis ($a_{\rm IMC}$). For an IMC positioned closer to the central SMBH, its secular resonance can extend significantly further inward, resulting in a smaller value of $a_{\rm sr, f}^{-}/a_{-}$. A similar trend is observed in the outer region, where a smaller $a_{\rm IMC}$ leads to a smaller value of $a_{+}/a_{\rm sr, f}^{+}$.

In the outer region of the IMC, the eccentricity of the IMC ($e_{\rm IMC}$) has a less pronounced impact on the value of $a_{+}/a_{\rm sr, f}^{+}$. For instance, IMCs with $e_{\rm IMC} = 0.4$ or $e_{\rm IMC} = 0.8$ yield comparable values of $a_{+}/a_{\rm sr, f}^{+}$. However, the size of the inner effective SR sweeping region is more sensitive to the IMC's eccentricity. Eccentricities that are either too small or too large result in a smaller effective region, with $a_{\rm sr, f}^{-}/a_{-}$ approaching unity. Only for IMCs with eccentricities in the range of $\sim 0.2-0.6$ and orbits within 0.1 pc do we observe a larger effective SR inner region, characterized by $a_{\rm sr, f}^{-}/a_{-} < 0.5$.

It is important to note that a larger effective SSR region does not necessarily correlate with a higher TDE enhancement rate. For example, in the fiducial model discussed above, with $a_{\rm IMC} = 0.5$ pc and $e_{\rm IMC} = 0.3$ (indicated by the red star in the figure), the effective secular resonance regions both inside and outside the IMC's orbit are relatively small, with values of $a_{\rm sr, f}^{-}/a_{-}$ and $a_{+}/a_{\rm sr, f}^{+}$ around $\sim 0.9-0.99$, close to unity. However, due to the short duration of the SR, which is within or approximately half of the depletion timescale, we can still expect a detectable TDE rate above $10^{-4}~{\rm yr}^{-1}$.

%\begin{figure*}
%\centering
%\gridline{\fig{figs/p_vtde_t_ai.pdf}{0.5\textwidth}{Semi-major axis of the IMC dependence}
%\fig{figs/p_vtde_t_ei.pdf}{0.5\textwidth}{Eccentricity of the IMC dependence}
%\fig{figs/p_vtde_t_tdep.pdf}{0.5\textwidth}{Depletion timescale of the gas disk dependence}
%\fig{figs/p_vtde_t_q.pdf}{0.5\textwidth}{Mass ratio of the IMC dependence}
%}
%\caption{The minimum mass ratio of the gas disk ($p_{\rm min}$, see Equation \ref{eq:pt0}) and the tidal disruption rate (Equation \ref{eq:v_ssr}) as function of time ($t/\tau_{\rm dep}$). The fiducial model adopt the IMC with the parameters: $a_{\rm I} = 0.01$ pc, $e_{\rm I} = 0.1$, $q = 10^{-3}$, and the gas disk with depletion timescale: $\tau_{\rm dep} = 1$ Myr. The dependence of those parameters ($a_{\rm I}$, $e_{\rm I}$, $\tau_{\rm dep}$ and $q$) are discussed with various colors, shown in the (a), (b), (c), (d) panels, respectively.}
%\label{fig:p_vtde_t}
%\end{figure*}
%

%In the figure \ref{fig:nssr}, we find that compared to the eccentricity of the IMC, the maximum TDE rate under the effect of the SSR mechanism during gas disk depletion depends more sensitively on the semi-major axis as well as the mass of the IMC. Generally, an outer IMC can trigger more stars excited to the tidal disruption zone on a short period as its secular resonance can sweep through in an extensive region, but this high TDE rate can only be maintained for a short period due to the more and more thin gas disk. And the SSR mechanism is less effective with a rapidly depleted gas disk or a less massive IMC. 

%
\begin{figure*}
%\centering
\gridline{\fig{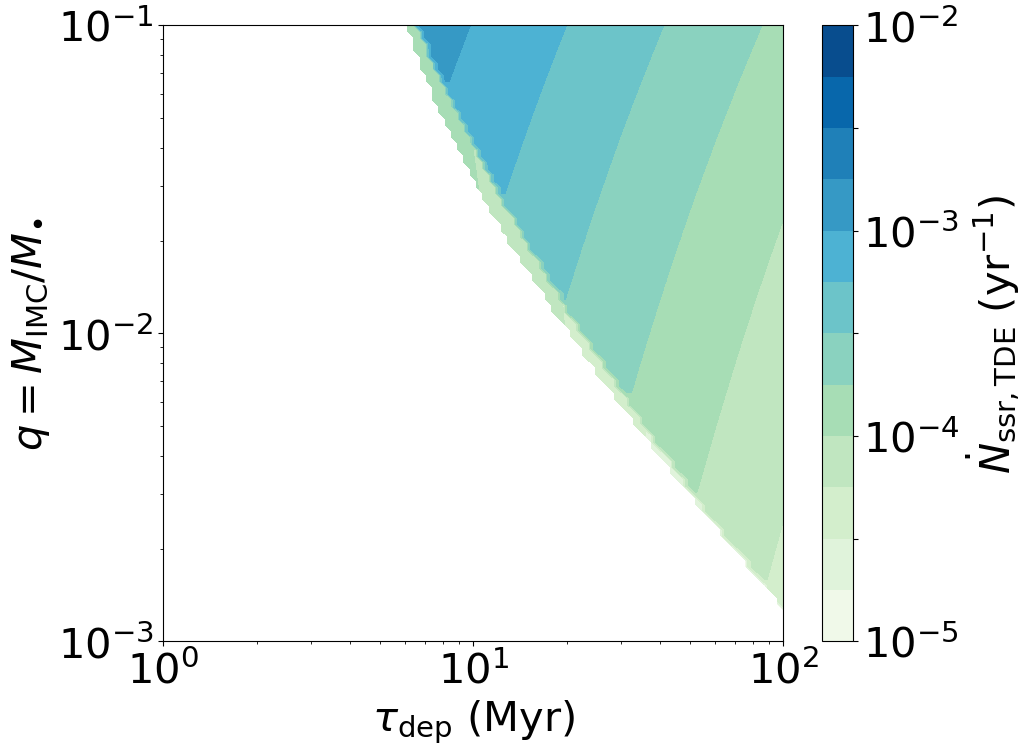}{0.5\textwidth}{(a) $a_{\rm IMC}$ = 1 pc, $e_{\rm IMC} = 0.3$}
\fig{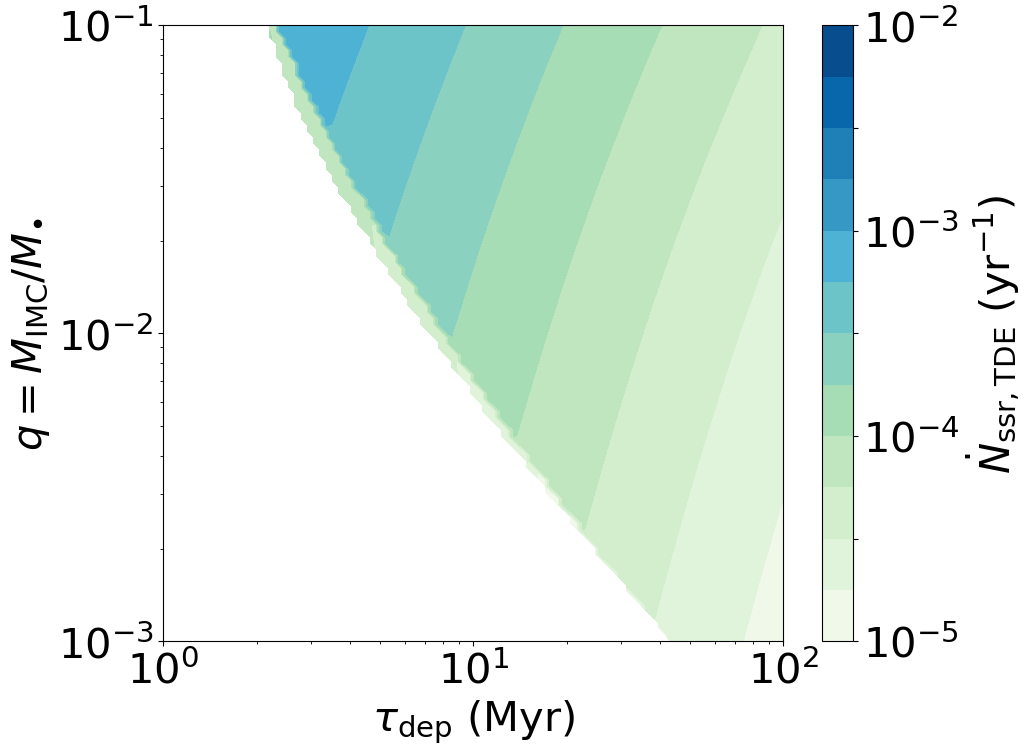}{0.5\textwidth}{(b) $a_{\rm IMC}$ = 0.5 pc, $e_{\rm IMC} = 0.3$}
}
\gridline{\fig{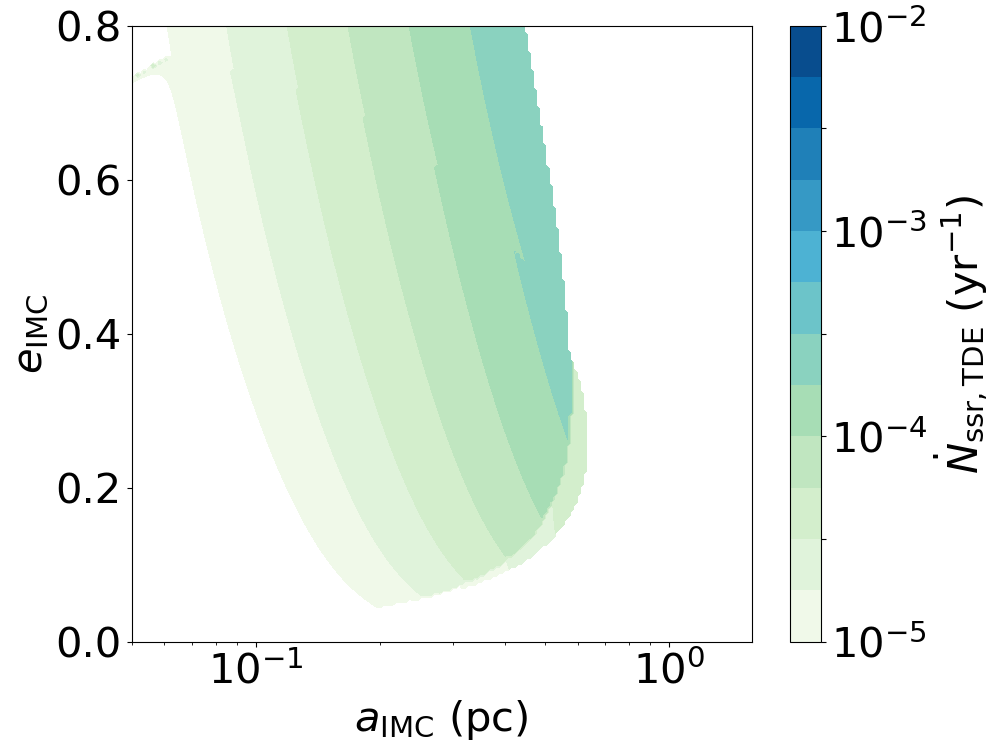}{0.5\textwidth}{(c) $q = 0.01$, $\tau_{\rm dep} = 10$ Myr}
\fig{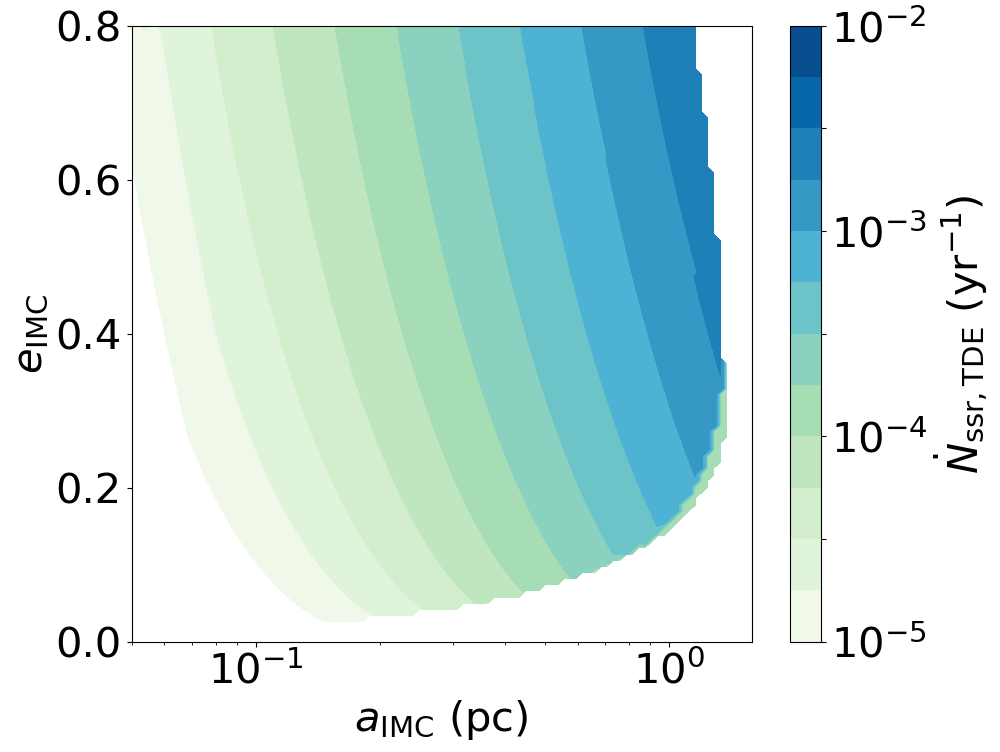}{0.5\textwidth}{(d) $q = 0.1$, $\tau_{\rm dep} = 10$ Myr}
}
\caption{{\bf Top:} The estimated TDE rate induced by the SSR mechanism (Equation~\ref{eq:v_ssr}) as a function of the IMC mass ratio $q = M_{\rm IMC}/M_{\bullet}$ and the gas disk depletion timescale $\tau_{\rm dep}$. {\bf Bottom:} The corresponding TDE rate as a function of the IMC's orbital semi-major axis $a_{\rm IMC}$ and eccentricity $e_{\rm IMC}$. The color bar indicates the TDE rate. 
%The fiducial model uses parameters $a_{\rm I} = 0.5$ pc, $e_{\rm I} = 0.3$, $q = 10^{-2}$, and $\tau_{\rm dep} = 10$ Myr. Each panel explores the dependence of the TDE rate on a single parameter.
%The IMC mass ratio, $q$, is set $0.01$. And the gas disk depletion timescale ($\tau_{\rm dep}$) is considered with 10 Myr. 
%Plotting the corresponding TDE rate as a function of the orbital parameters ($a_{\rm I}$ and $e_{\rm I}$), labeled with color bar.
}
\label{fig:nssr}
\end{figure*}

Figure \ref{fig:nssr} explores the parameter dependence of the TDE rate derived from Equation \eqref{eq:v_ssr}, focusing on the mass ratio $q \equiv M_{\rm IMC}/M_{\bullet}$, the disk depletion timescale $\tau_{\rm dep}$, and the orbital elements of the IMC, including the semi-major axis $a_{\rm IMC}$ and eccentricity $e_{\rm IMC}$. The models assume a central SMBH mass $M_{\bullet} = 10^7 M_{\odot}$, with a Schwarzchild radius $s = 10^{-6}$ pc, and a stellar cluster scale radius $R_{h} = 5.4$ pc. We consider only TDE rate enhancements exceeding $10^{-5}~{\rm yr}^{-1}$, which represents the baseline TDE rate for galactic nuclei driven by two-body relaxation processes.

The top row (panels a and b) characterizes the variation of the TDE rate enhancement with the accretion disk properties $\tau_{\rm dep}$ and $q$. Panel (a) corresponds to $a_{\rm IMC} = 1$ pc, while panel (b) adopts $a_{\rm IMC} = 0.5$ pc. Both assume a fiducial IMC eccentricity $e_{\rm IMC} = 0.3$. The bottom row (panels c and d) illustrates the sensitivity of the TDE rate to the IMC’s orbital configuration, with panel (c) for $q = 0.01$ and panel (d) for $q = 0.1$. 

Our analysis yields the following key findings:

\begin{itemize}

\item
The TDE enhancement driven by the SSR exhibits a strong, monotonic dependence on the accretion disk lifetime ($\tau_{\rm dep}$) before the peak, followed by an immediate drop. For short-lived disks ($\tau_{\rm dep} \lesssim$ a few Myr), the TDE rate remains low ($< 10^{-5} {\rm yr}^{-1}$) even in the presence of a massive IMC ($q \sim 0.1$), as the disk dissipates before it can efficiently perturb a significant number of stars. The peak TDE enhancement occurs for disks with intermediate lifetimes ($\tau_{\rm dep} \sim 10$ Myr), which provide sufficient time for secular perturbations to operate effectively. For instance, a system with $a_{\rm IMC} = 1$ pc and $e_{\rm IMC} = 0.3$ can achieve an enhancement of $\sim 10^{-3}~{\rm yr}^{-1}$ for $\tau_{\rm dep} = 10$ Myr. In contrast, for long-lived disks ($\tau_{\rm dep} \gtrsim$ tens of Myr), the peak TDE rate diminishes. Although the cumulative number of TDEs increases linearly with time, these events are distributed over a longer duration, thereby reducing the instantaneous flux.

\item
More massive IMCs (higher $q$) are more efficient at secularly perturbing stars into loss-cone orbits. A smaller IMC semi-major axis can elevate the TDE rate, but its influence is confined to a limited parameter space. A distant, massive IMC is most effective at triggering numerous TDEs. However, a close-in, less massive IMC ($q \sim 10^{-3}-10^{-2}$) can also induce observable TDE enhancements if paired with a long-lived gas disk ($\tau_{\rm dep} > 10$ Myr). For a typical AGN disk ($\tau_{\rm dep} \sim$ few Myr), an IMC with $a_{\rm IMC} = 0.5$ pc and $q = 0.05-0.1$ yields a maximum TDE rate enhancement of $\sim 10^{-3}~{\rm yr}^{-1}$.

\item
A higher IMC eccentricity increases the available angular momentum deficit, facilitating momentum transfer to cluster stars and driving them onto parabolic orbits. Consequently, the TDE rate increases modestly with $e_{\rm IMC}$, as shown in panels (c) and (d). However, high eccentricity also enlarges the gap region around the IMC (Equation \ref{eq:ainout}), which reduces the effective SSR-influenced region and thus the tidal disruption rate ${\dot N}_{\rm ssr, TDE}$ (Equation \ref{eq:v_ssr}). Overall, the TDE enhancement is less sensitive to $e_{\rm IMC}$ than to $a_{\rm IMC}$, $q$, and $\tau_{\rm dep}$. For a massive IMC ($q = 0.1$) with $\tau_{\rm dep} \sim 10$ Myr, an observable TDE enhancement rate can be induced across a broad range of orbital eccentricities.

\end{itemize}

This parameter study demonstrates that an IMC can significantly enhance the TDE rate in an AGN environment under a range of conditions. The most critical factors are the disk depletion timescale and the IMC mass and semi-major axis, while eccentricity plays a secondary role. We note that the estimated TDE rates should be regarded as conservative lower limits, as the sufficient condition for parabolic orbit excitation ($\tau_{\rm para}/\Delta \tau_{\rm sr} \leq 1$, Figure \ref{fig:a_tpara_sim}) suggests the number of excited stars may be higher than our baseline calculation.

%\section{Results}

%\subsection{Co-orbiting case}

%\begin{figure*}
%\centering
%\includegraphics[width=2\columnwidth]{figs/a0_rmin_com.pdf}
%\caption{The minimum distance of stars to the central SMBH as a function of their initial location (= periapse as all stars are initially in circular orbits) before they are ejected from the center ($e_{\star} > 1$) under the influence of the SSR mechanism. We set the distance of stars to be logarithmic distributed between $3 \times 10^{-4}$ and $9 \times 10^{-2}$ pc. The tidal disruption radius for $1 M_{\odot}$ is approximately $5 \times 10^{-6}$ pc as the SMBH is set with the mass $10^7 M_{\odot}$. We set the IMC mass to $10^4 M_{\odot}$ with various semi-major axes and eccentricities, range in $0.005-0.02$ pc and $0.1-0.5$, respectively. The crucial time for a star reached its closest approach to the central SMBH is labeled with colors.}
%\label{fig:aj_ej_com}
%\end{figure*}

\section{Summary and Discussions}
\label{sec:summary}

In this study, we have explored a mechanism, SSR, to explain the enhancement of TDE rates in AGNs. This mechanism is driven by an IMC and a depleting gaseous disk. Our model provides a detailed analysis of the gravitational interactions between stars, an IMC, and an evolving accretion disk in the vicinity of an SMBH. 

Our findings suggest that the presence of an IMC may substantially increase TDE rates via the SSR mechanism. This process engages with a dispersing gas disk and its accompanying swarm of embedded stars in the galactic center region. The ubiquity of accretion disks in AGNs indicates that this TDE enhancement mechanism could be widespread. In addition, observed high TDE rates in certain galaxies or certain types of galaxy may serve as an indirect indicator of a previously unseen IMCs on parsec (or sub-parsec) scales. The primary advantage of the SSR mechanism is that it can significantly and rapidly excite the surrounding stars on a timescale far shorter than those of two-body relaxation, efficiently driving stars to disruption.

We have theoretically estimated the effective parameter space for the SSR mechanism by constructing a gravitational potential that includes the GR effect of the central SMBH, the Galactic bulge, a depleting gas disk, a stellar cluster, and the IMC itself. These components induce competing precessions that dominate in different regions and are crucial for establishing the resonance. Our key findings are summarized as follows:

\begin{itemize}
\item 
The SSR mechanism occurs in galactic centers hosting stars and gaseous disks due to resonances between the precession rates of stars and an intermediate mass companion object. As a gaseous disk depletes over time, the locations of resonances sweep through semi-major axis space, leading to the SSR. 
\item The SSR can significantly excite stellar eccentricities in the resonance region, enhancing TDE rates by $1–3$ orders of magnitude. This is particularly effective in AGNs hosting IMCs with semi-major axes $a_{\rm IMC} > 0.1$ pc and for disk depletion timescales $\tau_{\rm dep}$ exceeding a few Myr (Figure \ref{fig:nssr}). The resonance operates on orbital timescales, bypassing the slow process of two-body relaxation (\S \ref{sec:ssr}), and is most efficient for stars orbiting near the IMC's location.

\item 
We have mapped the parameter space governing the SSR mechanism, which includes the IMC-to-SMBH mass ratio ($q$), the disk depletion timescale ($\tau_{\rm dep}$), and the IMC's orbital elements ($a_{\rm IMC}$, $e_{\rm IMC}$). Our results indicate that the TDE enhancement is most sensitive to the IMC's semi-major axis and mass.

\item 
The TDE rate enhancement is substantial but transient.
It peaks for wider-separation IMCs and longer-lived gas disks but decays as the disk dissipates. The mechanism becomes ineffective for rapidly depleting disks or low-mass IMCs. The condition for efficient eccentricity excitation, $\tau_{\rm para}/\Delta \tau_{\rm sr} \leq 1$ (Eq. \ref{eq:necessarypara}), is most readily met in gas-rich environments.
%where the disk-to-SMBH mass ratio is $p = M_{\rm Disk}/M_\bullet \geq 10^{-3} - 10^{-1}$.
The minimum initial gas disk mass $p_{\text{min}}$ is set by the orbital parameters and mass of the IMC. Above this threshold, the SSR-driven enhancement of TDE rates becomes effectively saturated and insensitive to the disk mass. Consequently, the results in Figures \ref{fig:nssr} and \ref{fig:a_e_pmin} capture the essential physics: once the disk is sufficiently massive to initiate secular resonance sweeping, the TDE rate depends primarily on the IMC’s orbit and mass, rather than on the precise value of the disk mass.

\item 
Our analytical predictions are corroborated by N-body simulations performed with the REBOUND code.
These simulations confirm that the SSR mechanism can drive stars onto near-parabolic orbits susceptible to tidal disruption. Our analytical calculation further underscores the critical role of initial conditions, particularly the potential intermediate mass companion and the gas depletion law, in setting the TDE rate.

\end{itemize}

The SSR mechanism provides a plausible explanation for observed correlations between TDE rates and host galaxy properties \citep{2024ApJ...961..211M, 2020SSRv..216...32F}.
The presence of an IMC alone leads to resonance, but not a SSR process without a depleting gaseous disk. Therefore, the requirement for a depleting disk implies that elevated TDE rates observed in certain galactic subtypes, such as post-starburst or ``green-valley" galaxies, might relate to previous AGN activity in these systems and depleting accretion disks. Secondly, the requirement for an IMC implies that our predicted parameter space aligns with the properties of these systems, suggesting that post-merger IMCs may be common in their nuclei.

The SSR mechanism can also efficiently excite a significant population of young stars to high eccentricities. These stars originate from a common birth disk that was co-aligned with the orbit of the potential IMC. Although only a small fraction are ejected onto hyperbolic orbits ($e_{\star} \geq 1$), a substantial number attain eccentricities high enough to reach the inner galactic region, such as the domain of the S-stars ($<0.05$ pc) in the Galactic Center (GC).

Under two-body stellar relaxation alone, the eccentricity excitation of young stars in the GC is highly inefficient. As demonstrated by previous studies \citep{Zheng2025}, two-body relaxation can only induce moderate eccentricity growth, yielding $\langle e_{\star}^2\rangle \sim 0.1-0.2$ over an O/B star lifetime ($\tau_{\star} \lesssim 10$ Myr) for stars within 0.1 pc, far too low to contribute significantly to TDE rates. Exciting these young massive stars to near-parabolic orbits ($e_\star \gtrsim 0.5$), necessary for loss-cone penetration, occurs on a timescale of $\tau_{e} \sim \sigma_{\star}^2/\dot{\sigma_{\star}}^2 \propto \langle e_{\star}^2\rangle^2 a_\star^2 $, which typically spans a few to tens of Gyr \citep{Zheng2025}, greatly exceeding stellar lifetimes.

In contrast, the SSR mechanism offers a far more efficient channel for eccentricity excitation. To illustrate, consider the Milky Way case, the S-stars orbiting Sgr A$^\star$ undergo resonant relaxation (RR) with an outer population of high-eccentricity perturbers, including the observed Clock-wise disk stars (CWSs) and a more vast, off-disk stellar population \citep{1996NewA....1..149R, 2006ApJ...645.1152H, yululin2007, 2011PhRvD..84b4032K, kocsis2015}. The latter can be pre-excited by mechanisms such as the Kozai-Lidov effect and SSR \citep{Zheng2025}. The RR timescale for eccentricity evolution is $\tau_{\rm RR, e} \sim 3 P_{\star} M_\bullet/M_\star \sim 10^{6-7}$ yr \citep{Murray1999}, while the corresponding precession timescale at a distance $r$ is $\tau_{\rm RR, \varpi} \sim \tau_{\rm RR, e} / [5 N_\star^{1/2}(<r)]$ \citep{yululin2007, kocsis2015}, allowing their angular momentum vectors to evolve toward an isotropic distribution. This RR process facilitates cumulative angular momentum exchange (without significant energy transfer) between the inner stellar population and the eccentric outer intruders, driving the orbital distribution toward isotropy and allowing inner stars to attain the high eccentricities required for TDEs.

However, generating such highly eccentric inner stars in the vicinity of the SMBH requires a source of angular momentum deficit (AMD) \citep{laskar1997}. This deficit can be supplied by the infalling AMD from nearby, highly-eccentric intruding stars, a process made attainable through the sweeping secular resonance driven by the IMC.

%For S-stars with sufficiently high eccentricity, $\tau_{\rm RR, \varpi} \leq \tau_\star$, allowing their orbital orientations to randomize, which may further contribute to tidal disruption event (TDE) rates.

Our assumption of coplanarity was adopted to isolate the gravitational effects of the disk, which constitute the primary focus of our study on secular resonant dynamics. While effects such as star–disk interactions and star–star relaxation are not considered here, they are necessary for a complete description of stellar evolution in an AGN disk. For instance, vector resonant relaxation among young stars can alter orbital inclinations and may thicken or warp the stellar disk on short timescales. As shown by \citet{KocsisTremaine2011}, the timescale for vector resonant relaxation to significantly modify a star's angular momentum vector can be comparable to or shorter than the secular timescales examined in this work, particularly within the inner 0.5 pc. This process could therefore replenish the population of stars with moderate inclinations relative to the gas disk, or conversely, disrupt the coherent disk structure required for the resonant eccentricity growth we describe. In such cases, the Kozai mechanism may become relevant. A quantitative investigation of tidal disruption event rates under the combined influence of secular dynamics and stellar relaxation will be pursued in a future numerical study.

In a related context, \citet{Xian2025} examined collisions between an orbiter, such as a main-sequence (MS) star, compact object (e.g., stellar-mass black hole, white dwarf), or a stripped stellar core following a partial tidal disruption event, and the accretion disk around a supermassive black hole with masses in the range $10^5–10^7 M_{\odot}$. Their results indicated that MS stars are ultimately destroyed due to progressive ablation of their outer envelopes during repeated disk crossings. However, in that scenario, stellar orbits are not confined to the plane of the gaseous disk, in contrast to the coplanar configuration adopted here. Thus, although such collision effects are relevant for inclined orbits, they do not conflict with the fundamental mechanism of sweeping secular resonances, which is predicated on the assumption of coplanarity.

Additionally, the present work neglects gas drag forces acting on highly eccentric stellar orbits during repeated disk passages, which could dissipate orbital energy and lead to circularization. However, \citet{MacLeod2020} demonstrated that such drag has a minimal impact on the overall rate of stellar tidal disruptions. Stars originating from more distant orbits, which receive sufficiently large per-orbit perturbations to be scattered over the disk loss cone, remain efficiently delivered to the black hole and undergo tidal disruption.

In this paper, we adopted a fixed orbit for the IMC as a first-order approximation. Generally speaking, a massive object moving through a sea of lighter stars experiences dynamical friction, which can lead to orbital decay. However, the regime changes once the IMBH forms a bound binary with the central SMBH. At this stage, the binary is energetically more bound than the typical surrounding stars and is classified as a ``hard'' binary. In this hard binary state, the evolution is no longer dominated by smooth dynamical friction from the stellar background but by three-body interactions, where stars are ejected via the gravitational slingshot mechanism, leading to binary hardening.
Crucially, the timescale for the initial inspiral due to dynamical friction can be extremely long for low mass-ratio binaries. As highlighted in the work of \cite{Dosopoulou2017}, which specifically models the dynamical friction evolution of a secondary black hole in a stellar cusp around a primary SMBH, secondary holes with mass ratios $q \lesssim 10^{-3}$ in massive galaxies can have decay timescales exceeding a Hubble time. Their work suggests that the IMBH would stall at a distance on the order of the primary's influence radius (tens to hundreds of parsecs) for many Gyr. In the context of our work, both the mass ratio and the primary mass of the SMBH span a broad range. Consequently, the ``hard" separation between the SMBH and IMC correspondingly spans a large range, from sub-parsec to a few parsecs.
Furthermore, as reviewed in \cite{Kelly2017}, the subsequent hardening phase for such a hard binary is also a very slow process, often requiring many hundreds of Myr to reduce the separation significantly. Therefore, for the relatively short timescales considered in our study (the disk depletion and SR sweeping timescale of $\sim$ 10 Myr), the assumption of a fixed, non-decaying orbit for the IMC is a reasonable and justifiable approximation. The expected orbital decay from dynamical friction is negligible over this period.
However, we acknowledge that \cite{Kelly2017} also identified a strong trend towards shorter lifetimes for binaries with simultaneously high total masses and mass ratios. If, in some regions of our parameter space, the dynamical fraction timescale were much shorter than the gaseous disk depletion timescale, the necessary condition for the SSR mechanism to trigger an enhanced TDE rate would no longer hold. During the inspiral of the IMC induced by dynamical friction, however, many stars would be captured by the IMC's mean motion resonances and experience eccentricity growth, which may subsequently lead to tidal disruption events. We plan to explore this related effect in a follow-up paper.

In conclusion, the SSR mechanism offers a robust, time-dependent framework for enhancing TDE rates. Future high-cadence monitoring of AGN could test our model by searching for TDE flares correlated with kinematic signatures of IMCs or asymmetric disk structures. Our work underscores the dynamic interplay between stellar clusters, accretion disks, and SMBHs in driving extreme astrophysical phenomena.

%\nolinenumbers

%\begin{acknowledgments}
\section*{Acknowledgments}

Simulations in this paper made use of the REBOUND code which is publicly available at \href{http://github.com/hannorein/rebound}{REBOUND} website.  
The authors acknowledge the Tsinghua Astrophysics High-Performance Computing platform at Tsinghua University for providing computational and data storage resources that have contributed to the research results reported within this paper.  
This work is supported by the China Manned Space Program
with grant no.CMS-CSST-2025-A16.
XCZ is supported by the National Natural Science Foundation of China (Grant No.12203007) and the Mengya Program of Beijing Academy of Science and Technology (BGS202203). ZZS is supported by the National Natural Science Foundation of China (Grant No.12503027). YY is supported by the Young Data Scientist Program of the China National Astronomical Data Center (No. NADC2025YDS-02).
%\end{acknowledgments}

%%%%%%%%%%%%%%%%%%%% REFERENCES %%%%%%%%%%%%%%%%%%

\bibliography{tde.bib}{}
\bibliographystyle{aasjournal}

%%%%%%%%%%%%%%%%%%%%%%%%%%%%%%%%%%%%%%%%%%%%%%%%%%

%%%%%%%%%%%%%%%%%%%%%%%%%%%%%%%%%%%%%%%%%%%%%%%%%%

\vspace{15mm}

\software{REBOUND \citep{reinliu2012, rein2019}}

\vspace{15mm}

% Don't change these lines
%\bsp	% typesetting comment
%\label{lastpage}
\end{CJK*}
\end{document}